\documentclass[fleqn,10pt]{wlscirep}
\usepackage[utf8]{inputenc}
\usepackage[T1]{fontenc}
\usepackage{bm}
\usepackage{color}
\usepackage{enumitem}
\usepackage{hyperref}
\usepackage{verbatim}
\usepackage{amsmath,amstext,mathrsfs}
\usepackage[all]{hypcap} 
\usepackage{afterpage}    
\usepackage{marginnote}
\usepackage{makecell}
\usepackage{subfigure}
\usepackage{multirow}
\usepackage{lineno}
\usepackage{wrapfig}
\usepackage[para]{footmisc}
\usepackage{float}
\title{\LARGE{Role of magnetic fields in fueling Seyfert nuclei}}

\author[1,2,*]{Yue Hu}
\author[2,3,*]{Alex Lazarian}
\author[4]{Rainer Beck}
\author[5]{Siyao Xu}
\affil[1]{\small{Department of Physics, University of Wisconsin-Madison, Madison, WI 53706, USA}}
\affil[2]{\small{Department of Astronomy, University of Wisconsin-Madison, Madison, WI 53706, USA}}
\affil[3]{\small{Centro de Investigación en Astronomia, Universidad Bernardo O’Higgins, Santiago, General Gana 1760, 8370993, Chile}}
\affil[4]{\small{Max-Planck-Institut fur Radioastronomie, Auf dem Hugel 69, 53121 Bonn, Germany}}
\affil[5]{\small{Institute for Advanced Study, 1 Einstein Drive, Princeton, NJ 08540, USA (Hubble Fellow)}}
\affil[*]{\small{e-mail: yue.hu@wisc.edu; lazarian@astro.wisc.edu}}

\begin{abstract}
Molecular gas is believed to be the fuel for star formation and nuclear activity in Seyfert galaxies. To explore the role of magnetic fields in funneling molecular gas into the nuclear region, measurements of the magnetic fields embedded in molecular gas are needed. By applying the new velocity gradient technique (VGT) to ALMA and PAWS's CO isotopolog data, we obtain the first detection of CO-associated magnetic fields in several nearby Seyfert galaxies and their unprecedented high-resolution magnetic field maps. The VGT-measured magnetic fields in molecular gas globally agree with those inferred from existing HAWC+ dust polarization and VLA synchrotron polarization. An overall good alignment between the magnetic fields traced by VGT-CO and by synchrotron polarization may support the correlation between star formation and cosmic ray generation. We find that the magnetic fields traced by VGT-CO have a significant radial component in the central regions of most Seyferts in our sample, where efficient molecular gas inflows or outflow may happen. In particular, we find local misalignment between the magnetic fields traced by CO and dust polarization within the nuclear ring of NGC 1097, and the former aligns with the central bar's orientation. This misalignment reveals different magnetic field configurations in different gas phases and may provide an observational diagnostic for the ongoing multi-phase fueling of Seyfert activity.
\end{abstract}
\begin{document}

\flushbottom
\maketitle
\thispagestyle{empty}
\vspace{-0.5cm}
Magnetic fields, as the major component of interstellar media (ISM) \cite{2005A&A...444..739B,2013pss5.book..641B}, impact star formation \cite{MO07}, gas flows in spiral arms and bars \cite{2020NatAs...4..704A}, and the formation and evolution of spiral galaxies. The study of magnetic fields in one of the two classes of active galaxies, Seyfert galaxies, is essential for identifying their role in funneling the circumnuclear molecular gas, which holds the key for understanding the energy source of Seyferts and the connection between the enhanced star formation and Seyfert activity \cite{Maio97}. However, progress is hampered by a lack of measurements of the magnetic fields directly associated with (extragalactic) molecular gas, which is the potential fuel powering the nuclear starburst and Seyfert activities.

The plane-of-the-sky (POS) magnetic field distribution is traditionally measured using polarization. The synchrotron polarization \cite{2005A&A...444..739B,2011MNRAS.412.2396F} has recently been augmented by the measurements of polarized dust emission by the Stratospheric Observatory for Infrared Astronomy (SOFIA) \cite{2020ApJ...888...66L,2021arXiv210709063L,M51}. It is believed that polarization from dust is weighted with gas density and thus it is biased toward the magnetic fields in high-density media, while the synchrotron polarization depends on the acceleration and propagation of cosmic-ray electrons and samples the magnetic fields in more diffuse media. These two approaches have their limitations. For instance, dust polarization is sensitive to the dust alignment efficiency \cite{2015ARA&A..53..501A}, and synchrotron polarization can be distorted by Faraday rotation \cite{LP16}.  More importantly, neither of them can separate out the magnetic fields associated with molecular gas from multiphase ISM.

In this paper, we employ a novel technique, the Velocity Gradient Technique (VGT\cite{2017ApJ...835...41G,YL17a,LY18a,PCA}), to exclusively trace the magnetic fields in molecular gas. As CO is the best tracer of nuclear gas dynamics \cite{Garc03}, the application of VGT to the high-resolution CO maps of nearby Seyfert galaxies holds the promise in understanding the role of magnetic fields in transporting the potential fuel towards the nuclei and the origin of nuclear activity. The VGT is established based on both observations showing the ubiquity of turbulence \cite{2010ApJ...710..853C} and the recent progress in understanding the fundamental effects of magnetic fields on turbulence \cite{GS95,LV99}. As demonstrated by both state-of-the-art magnetohydrodynamic (MHD) simulations and in-situ measurements in the solar wind\cite{2002ApJ...564..291C,2021ApJ...915L...8D}, turbulent motions in the presence of magnetic fields are preferentially in the direction perpendicular to the magnetic field due to the minimum resistance in this direction. As a result, the gradients of turbulent velocities are perpendicular to the local magnetic field. Therefore, by measuring the velocity gradient of molecular gas' turbulent motions, one obtains maps of POS magnetic fields in the molecular gas phase. 

The VGT has been extensively tested by numerical simulations in various astrophysical conditions \cite{LY18a,H2,2021ApJ...911...53H}, considering CO self-absorption \cite{2019MNRAS.483.1287G,2019ApJ...873...16H,2021MNRAS.502.1768H}, self-gravity effects \cite{doublepeak},  and recently applied to mapping magnetic fields in the Milky Way in different phases, from diffuse H I media \cite{YL17a,2020ApJ...888...96H,2020MNRAS.496.2868L,2020RNAAS...4..105H} to dense molecular clouds \cite{survey,2021ApJ...912....2H,2020arXiv200715344A,2021arXiv210913670L}, as well as the Central Molecular Zone \cite{CMZ,2022MNRAS.tmp.1055H}. The requirement for the application of the VGT is that the turbulent injection scale is resolved, which is $\sim 100$ pc for the Milky Way \cite{2010ApJ...710..853C} and expected to be similar in other spiral galaxies. Therefore, with high-resolution ALMA \cite{2017PASJ...69...18T,2021arXiv210407739L} and PdBI Arcsecond Whirlpool Survey (PAWS\cite{2013ApJ...779...46H}) spectroscopic data, the VGT opens a radically new avenue to mapping magnetic fields of nearby galaxies that is complementary to both synchrotron and dust polarization measurements. 

In this work, we apply the VGT to five nearby Seyfert galaxies, M51\cite{2013ApJ...779...46H}, NGC 1068 \cite{2017PASJ...69...18T}, NGC 1097 \cite{2021arXiv210709063L}, NGC 3627 \cite{Soi01}, and NGC 4826 \cite{Garc03}, by using molecular emission line data of CO isotopologs obtained with ALMA and PAWS archives \cite{2013ApJ...779...46H,2017PASJ...69...18T,2021arXiv210407739L,2021ApJS..255...19L}. We compare our results of the first four galaxies with the dust polarization data from the SOFIA legacy program \cite{2020ApJ...888...66L,M51,2021arXiv210709063L} and synchrotron polarization data from the Very Large Array (VLA) \cite{2005A&A...444..739B,2011MNRAS.412.2396F}. We demonstrate that the synergy of the VGT using gas tracers and polarization presents valuable information on magnetic fields in different gas phases and their role in fueling nuclear activity in Seyferts. We also use the VGT to produce the first magnetic field map of NGC 4826, for which no polarization data are available yet. 

\begin{figure}[h!]
\centering
\includegraphics[width=1.0\linewidth]{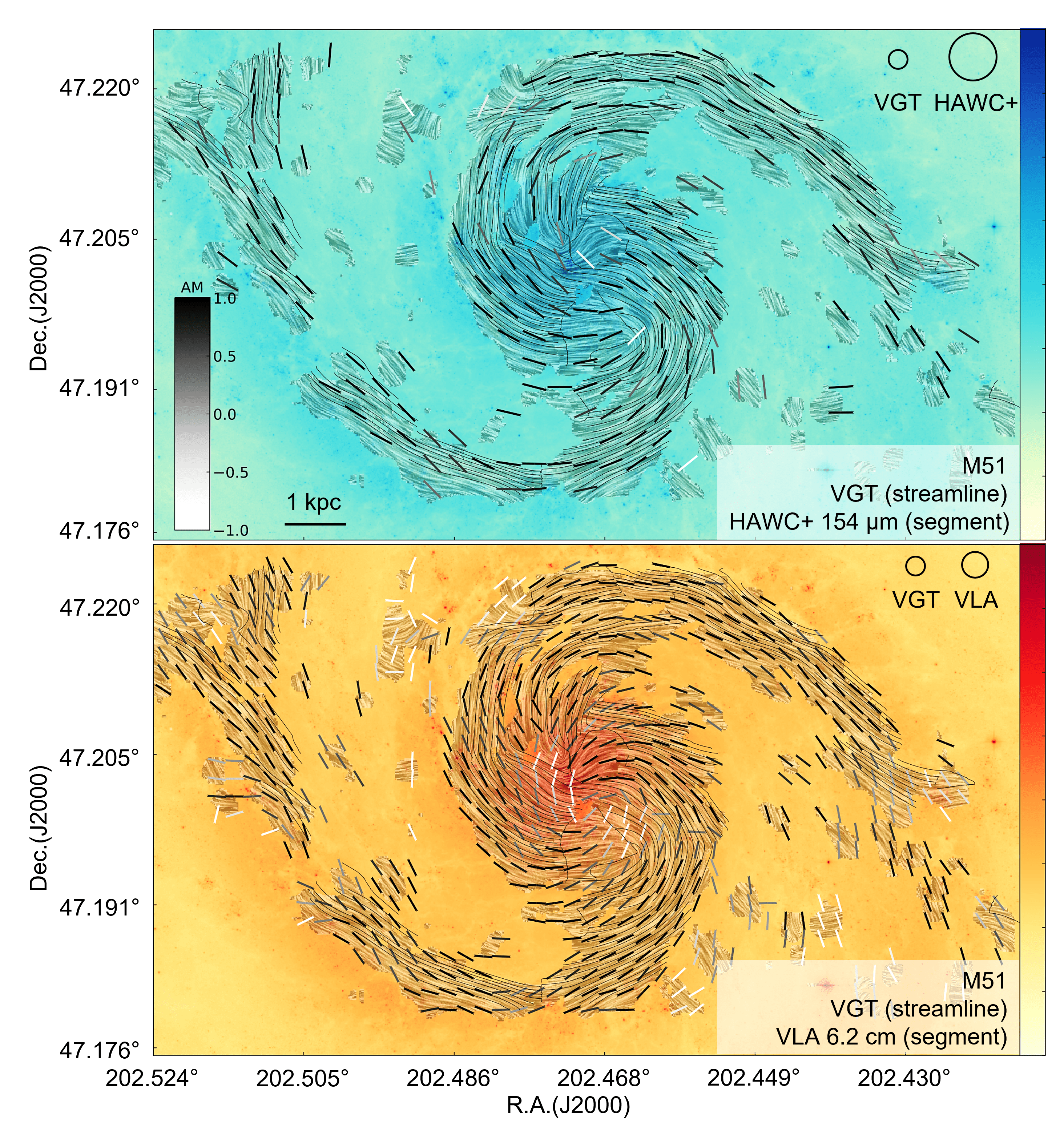}
\caption{Differences and similarities of magnetic fields in different gas phases towards the M51 galaxy. \textbf{Top:} morphology of magnetic fields revealed by the VGT using $\rm ^{12}CO$ (J = 1-0) emission line and HAWC+ polarization at 154 $\mu$m \cite{M51}. The VGT-measurement is visualized by black streamlines and HAWC+ is represented by the colored segments. Colors on polarization vectors present the AM of the VGT and polarization. The magnetic field is overlaid with the Hubble Space Telescope (HST) WFC3/F814W ultraviolet image. The black circle represents the beamwidth of observation. \textbf{Bottom:} Same as the top panel, but for the morphology of magnetic field, mostly in warm gas, revealed by VLA + Effelsberg polarization at 6.2 cm \cite{2011MNRAS.412.2396F} (colored segments) and VGT (black streamlines). The colorbars of background HST images are logarithmically spaced in the range from $10^{-2}$ to $10^2$ electrons per second. }
\label{m51}
\end{figure}

\begin{figure}[t]
\centering
\includegraphics[width=1.0\linewidth]{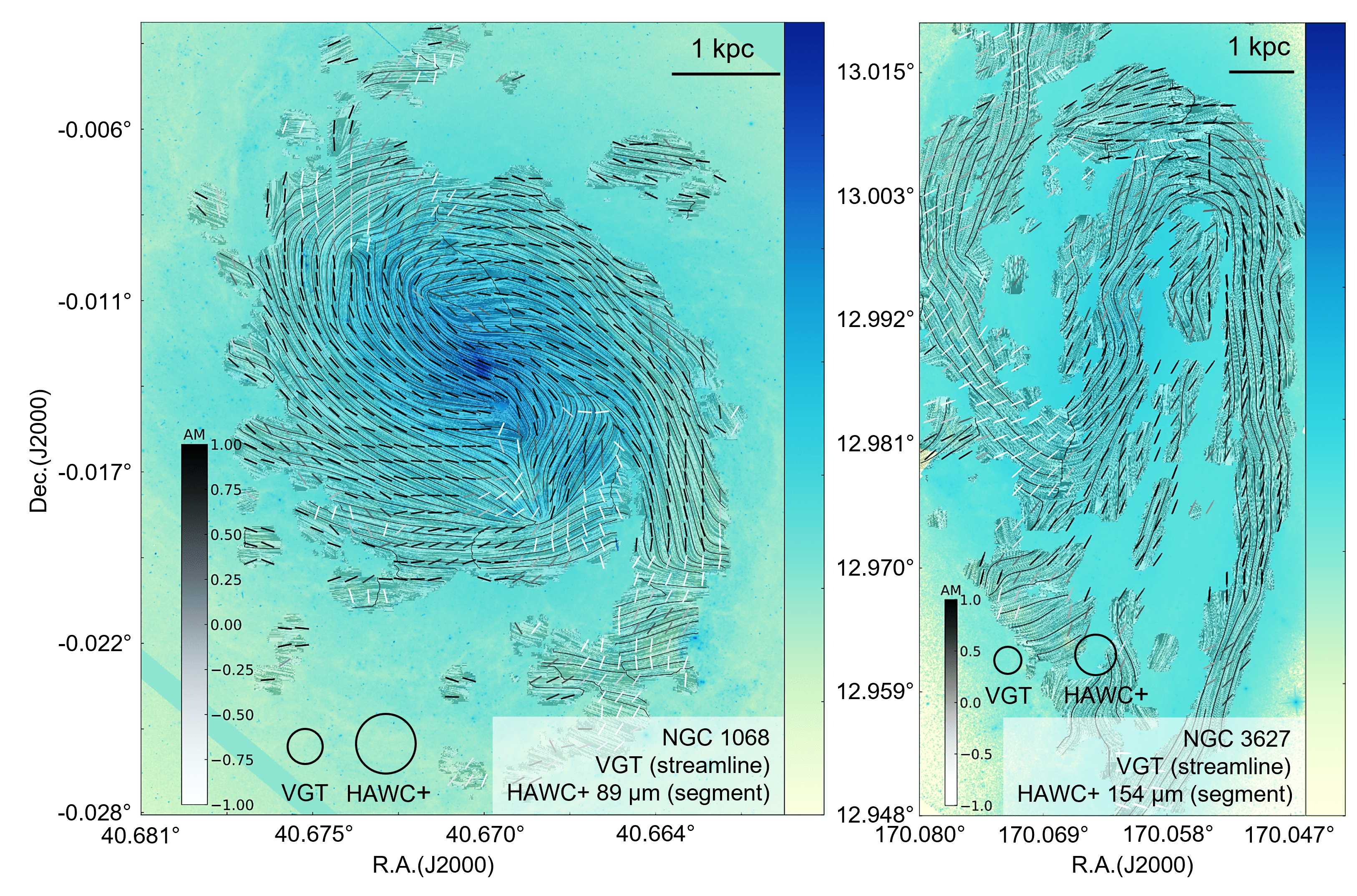}
\caption{Same as Fig.~\ref{m51}, but for the NGC 1068 (left) and NGC 3627 (right) galaxies. \textbf{NGC 1068:} the VGT uses $\rm ^{13}CO$ (J = 1-0) emission line (black streamlines) and HAWC+ polarization is observed at 89 $\mu$m (colored segments)\cite{2020ApJ...888...66L}. The colorbar of background HST WFC3/F814W ultraviolet image is logarithmically spaced in the range from $10^{-1}$ to $10^{2.5}$ electrons per second. \textbf{NGC 3627:} the VGT uses $\rm ^{13}CO$ (J = 1-0) emission line (black streamlines) and the HAWC+ 154 $\mu$m polarization (colored segments). The magnetic fields are overlaid with the HST WFC3/F814W ultraviolet images \cite{2022ApJS..258...10L}. The colorbar of HST image is logarithmically spaced in the range from $10^{-6}$ to $10^{6}$ electrons per second.}
\label{ngc1068}
\end{figure}

\begin{figure}[h]
\centering
\includegraphics[width=1.00\linewidth]{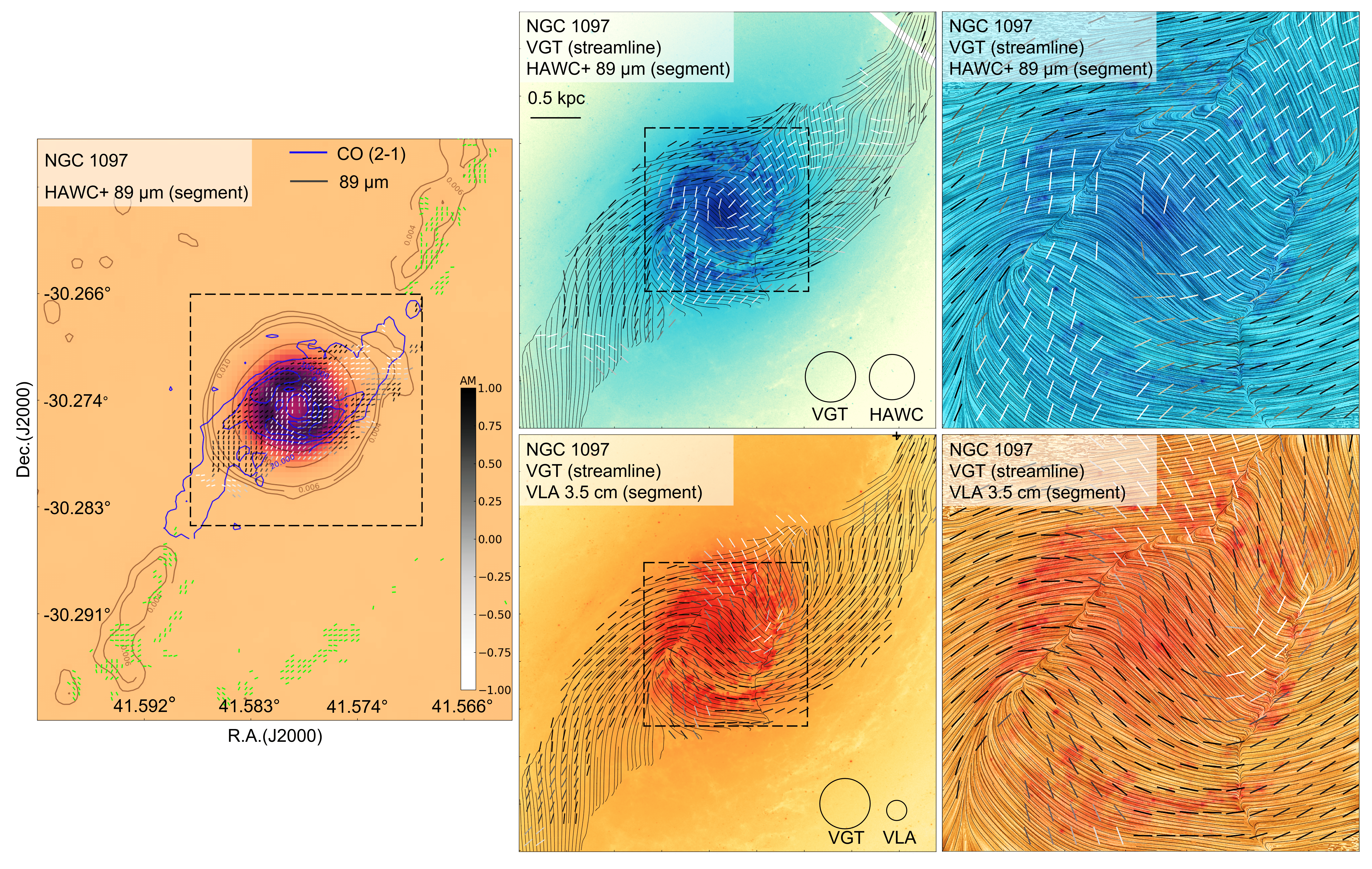}
\caption{\textbf{Left:} Morphology of magnetic fields revealed by the HAWC+ polarization at 89 $\rm{\mu}$m towards NGC 1097. The magnetic field is overlaid with the 89 $\mu$m continuum intensity map. Colors, except the green, on polarization vectors represent AM of the VGT and polarization. Dark grey contours represent the dust lanes and the central ring observed at 89 $\mu$m, while dark blue contours mean CO structures. The colorbar of the 89 $\mu$m continuum image is uniformly spaced in the range from $0$ to $0.45$ Jy per pixel. \textbf{Top left:} A zoom-in view of the magnetic fields in NGC 1097's circumnuclear  region,  mapped with the VGT using $\rm ^{12}CO$ (J = 2-1) emission (black streamlines)  and HAWC+ polarization at 89 $\mu m$ \cite{2021arXiv210709063L} (colored segments). The magnetic fields are overlaid onto the HST WFC3/F814W ultraviolet image \cite{2022ApJS..258...10L}. \textbf{Top right:} a further zoom-in on the magnetic fields in the starburst ring and the inner bar. \textbf{Bottom left and right}: same as the top panel, but for the VGT (black streamlines) and VLA + Effelsberg polarization (colored segments) observed at 3.5 cm \cite{2005A&A...444..739B}. The colorbars of background HST images are logarithmically spaced in the range from $10^{-2}$ to $10^2$ electrons per second.}
\label{ngc1097}
\end{figure}

\begin{figure}[t]
\centering
\includegraphics[width=0.88\linewidth]{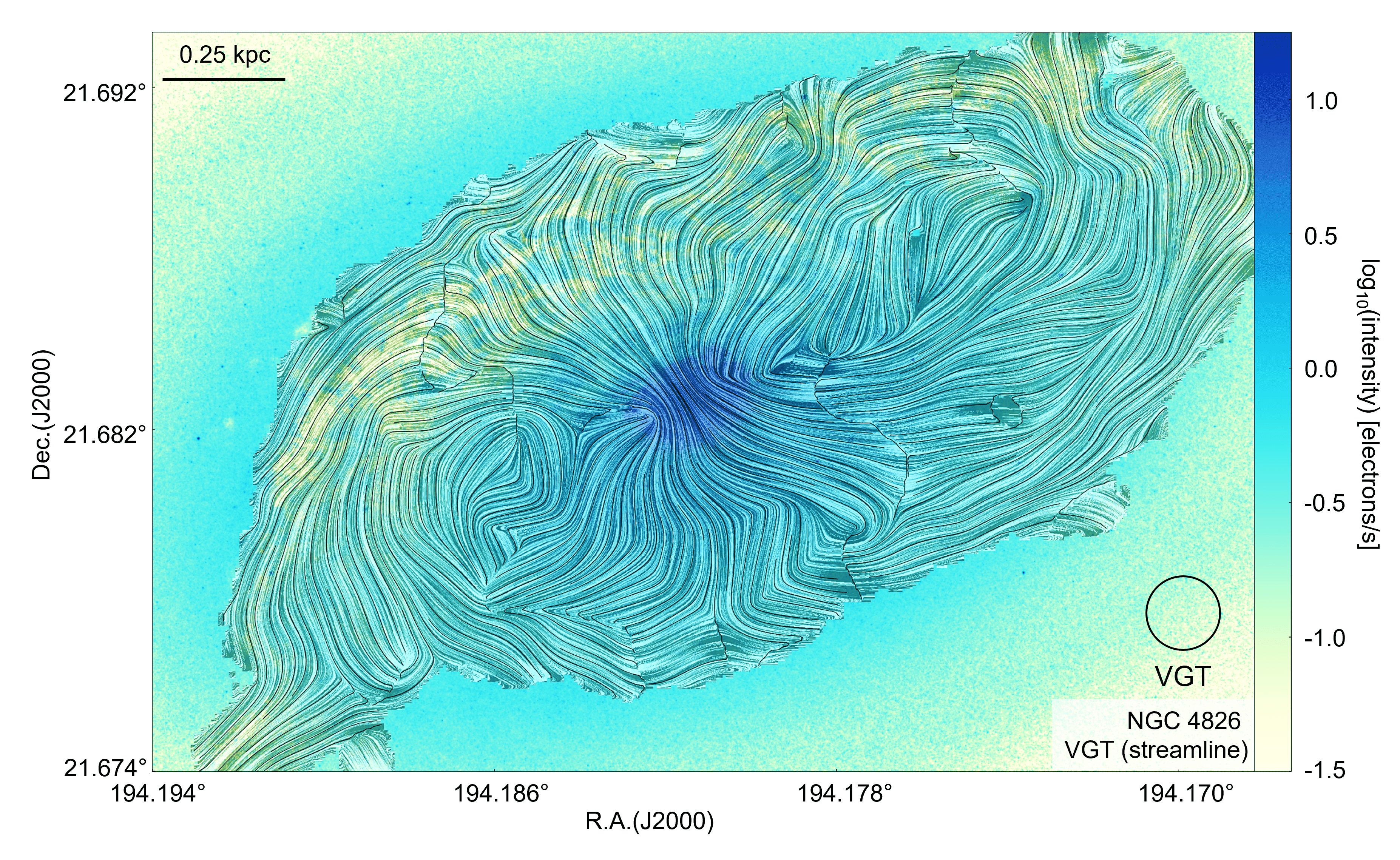}
\caption{ Morphology of magnetic fields revealed by the VGT (streamlines) using $\rm ^{12}CO$ (J = 2-1) emission line towards the NGC 4826 galaxy. The magnetic field is overlaid with the Hubble Space Telescope WFC3/F814W ultraviolet image \cite{2022ApJS..258...10L}. The white circle represents the beam width of observation.}
\label{ngc4826}
\end{figure}

\section*{Results}
\label{sec.result}
%\textbf{Alignment of magnetic fields traced by CO, dust, and synchrotron:} 
The input for the VGT is the high-resolution spectroscopic cubes of molecular tracers $\rm ^{12}CO$ and $\rm ^{13}CO$ (see the Method and the Supplementary Tab.~1 for details of the spectroscopic data). We follow the established VGT procedures \cite{CMZ} to trace the POS magnetic fields. Estimates on the statistical uncertainties of VGT measurements are provided in the Supplementary Fig.~\ref{error1}. We compare our magnetic field maps to those inferred from polarization obtained with SOFIA's High-resolution Airborne Wideband Camera Plus (HAWC+) as well as the VLA. The correlation between cosmic ray generation and star formation is suggested by the ``far-infrared-radio correlation" \cite{Matt21}. Thus radio synchrotron radiation is expected around star-forming regions, but with a much larger scale height due to the cross-phase diffusion of turbulent magnetic fields and diffusion of cosmic ray electron \cite{Planck14,XL22}. Dust polarization is expected to trace well the magnetic fields in cold dense media according to Planck results for the Milky Way \cite{Planck14}. 
%The deviations of the polarization from projected magnetic field can be due to the differences in grain alignment that can arise from the differences with the grain composition and environment in parts of Seyfert galaxies \cite{2015ARA&A..53..501A}.   
%{\sy (A comparison between their resolutions is better provided here.)}

The correspondence of the POS magnetic field orientations obtained with the VGT and polarization measurements is quantified by the \textbf{Alignment Measure} (AM\cite{2017ApJ...835...41G}): ${\rm AM}=2(\cos^{2} \theta_{r}-\frac{1}{2})$, where $\theta_r$ is the relative angle between the two magnetic field vectors. If the two measures provide identical results, there is $\rm AM = 1$. If the two measures are perpendicular to each, i.e., misaligned, we have $\rm AM = -1$.

\textbf{M51}:
Fig.~\ref{m51} presents the morphology of magnetic fields towards M51 measured by the VGT-$\rm ^{12}CO$, HAWC+ 154 $\mu$m dust polarization \cite{M51}, and VLA radio 6.2 cm polarization \cite{2011MNRAS.412.2396F}. For the VGT measurement, we average the gradients over each 20$\times$20 pixels sub-block, which is an optimal block size used in earlier VGT studies \cite{H2}, and smooth the gradients map $\psi_g$ with a Gaussian filter 
Full Width at Half Maximum (FWHM) $\sim6''$ (see the Method). The resulting magnetic field map has a higher resolution than that of HAWC+ ($\sim 13.6''$) and VLA ($\sim 8''$). From Fig.~\ref{m51}, we see that in spiral arms, the global magnetic structure traced by the VGT and polarization are in good agreement, with the inferred magnetic fields closely following the spiral arm pattern. On average, the AMs of the two methods are in the range of $0.75-1.00$ (Fig.~\ref{AM_hist} in the Supplementary), with a peak value at $\rm AM\approx 1$. In particular, we observe a better alignment in the regions of active star formation (Fig.~\ref{SFR} in the Supplementary). Noticeably, the misalignment (i.e., negative AM) between VGT and VLA measurements appears in the central region. However, as shown in the Supplementary Information, the VGT measurement in low-CO-intensity regions typically is associated with high uncertainty due to the poor signal-to-noise ratio. Such misalignment in low-intensity regions may not be caused by physical reasons. A radial component of magnetic fields is only significant in CO (Fig.~\ref{PM} in the Supplementary), accompanying the molecular gas inflow driven by the two-armed spiral towards the nucleus of M51 \cite{Quer16}.

%To what extend this misalignment can be attributed to the statistical uncertainties of the technique involved we analyze in the Supplementary material, see the Supplementary Fig.~{\color{blue} 6}.

\textbf{NGC 1068}:
In the Seyfert and starburst galaxy NGC 1068, molecular gas is abundant along the star-forming spiral arms and the central region. Fig.~\ref{ngc1068} shows the POS magnetic fields towards NGC 1068 mapped by applying the VGT (FWHM $\sim5''$) to $\rm ^{13}CO$ (J = 1-0) emission, in comparison with HAWC+ 89 $\mu$m polarization (beam size $\sim7.8''$). The two measurements are globally compatible with each other, except in the interface regions between the $\sim2.6$ kpc bar \cite{garca14} and the spiral arms. In the interface regions, strong inward radial flow of molecular gas is also detected and attributed to the combined action of the bar and the spiral \cite{garca14}. Beginning from the bar-arm transition regions, VGT reveals a significant radial component of magnetic fields in CO along the bar (Fig.~\ref{PM} in the Supplementary). The remarkable coincidence between the inflow and radial magnetic fields seen in molecular gas suggests that magnetic fields play an important role in transporting CO to the central reservoir of molecular gas and powering star formation in the kpc-scale starburst ring. However, the readers should note that in this work, we do not explicitly distinguish inflow or outflow. For a region dominated by gravity, we expect to observe accretion and inflows associated with molecular material. 

\textbf{NGC 3627}: 
For this interacting galaxy \cite{Hay79}, the magnetic fields traced by $\rm ^{12}CO$ (J = 2-1) closely follow the bar and asymmetric spiral structures, with FWHM$\sim8.0''$ (see Fig. \ref{ngc1068}). By comparing with HAWC+ 154 $\mu m$ polarization (beam size $\sim13.6''$), we see a good alignment between the two measurements in the western arm and the bar. The misalignment appears in the southern bar end and the eastern arm. An unusual magnetic field component crossing the dust lane in the southeast (SE) disk was earlier found by radio polarization measurements \cite{Soi01}. A recent collision with a dwarf galaxy may be responsible for the enhanced star formation in the eastern arm and the distortion of the SE region \cite{Wezg12}. The different magnetic field configurations seen in CO and dust support this scenario and may reflect distinct pre- and post-collision flows of different phases induced by the interaction. The tidal interaction with the neighbor
galaxy NGC 3628 can significantly affect its subsequent dynamical evolution and trigger efficient radial inflow toward the nucleus \cite{Zhang93}.

\textbf{Circumnuclear region of NGC 1097:}\\
\textbf{(1) A molecular ring in the nucleus and dust lanes in the bar:} As an efficient driver of gas inflow, a strong bar with prominent dust lanes along the bar is present in NGC 1097. The orbiting gas loses angular momentum at the dust-lane shocks and falls toward the galaxy center \cite{Atha92}. Despite the high density of gas due to shock compression, the formation of molecular clouds and stars are suppressed in the bar because of the strong shear along the bar \cite{Atha92}. The inflowing gas settles in the nuclear ring (see Fig.~\ref{ngc1097}), serving as the raw material for molecular cloud and star formation when the velocity shear is mitigated. In addition to the large shear along the dust lane, another possible cause of the different distribution between dust and molecular gas is temperature. When temperature increases, the shock front along the bar moves closer to the bar’s major axis, and the central ring that connects the inner ends of the bar becomes more elongated along the bar’s major axis \cite{EngP97,Pata00}. It means that the colder molecular gas traced by CO (and dust) and the warmer gas traced by dust can have different distributions and flow configurations. 

\noindent{\bf (2) Magnetic fields traced by CO:} 
As shown in Fig.~\ref{ngc1097}, at the inner ends of the bar, the magnetic fields mapped with CO (VGT; FWHM$\sim10''$) are curved into the central circumnuclear ring. Magnetic fields are compressed by shocks and stretched by shear in the dust lanes along the bar. The tension force of bent magnetic fields causes further removal of the angular momentum of gas both at dust-lane shocks and within the nuclear ring. MHD simulations of barred galaxies suggest that the presence of magnetic fields leads to a more centrally concentrated ring and enhanced mass inflow rate to the galaxy center\cite{KJ12}. As an important characteristic for one to identify the role of magnetic fields, we clearly see the magnetic fields mapped with CO are bent into an $''$L$''$ shape within the nuclear ring as expected from simulations \cite{KJ12}. We also see that the magnetic fields within the ring have a radial component (Fig.~\ref{PM} in the Supplementary) following the secondary bar of NGC 1097 reported by the ref.~\cite{Quil95}. The magnetic fields threading the secondary bar impose a magnetic braking effect on the gas spiraling into the innermost region and further remove its angular momentum, as suggested in the ref.~\cite{KJ12}. Our observed magnetic field morphology in CO suggests that in addition to the gravitational torques from the bars, magnetic fields introduce additional torques, which contribute to the removal of angular momentum and transport of gas from the galactic disk to its innermost region. 

\noindent{\bf (3) Magnetic fields traced by dust:} We see from Fig.~\ref{ngc1097} that the magnetic fields measured by 89$\mu m$ HAWC+ dust polarization ($\sim7.8''$) \cite{2021arXiv210709063L} extend into the ring. Unlike the magnetic fields traced by CO, there is no significant bending of field lines within the ring. There is a clear correlation between the regions with misalignment and radial field (Fig.~\ref{ampm} in the Supplementary), showing the radial magnetic fields across the ring along the primary large-scale bar seen in dust polarization. The misalignment between the magnetic fields traced by CO and dust can be attributed to their different distributions in the strong bar and inner ring system. If there is a warmer phase that is preferentially traced by dust, the morphology of magnetic fields with AM$\approx-1$ reflects the more elongated shape of the central concentration of warmer gas compared to that of colder gas. This misalignment suggests the two-phase gas inflows along the bar and within the ring. 
    
\noindent{\bf (4) Magnetic fields traced by synchrotron:} In Fig.~\ref{ngc1097}, we see an overall good alignment between the magnetic fields traced by CO and synchrotron. The molecular ring of NGC 1097 coincides with the starburst ring. This supports the correlation between star formation and cosmic ray generation. Particularly, the ref.~\cite{2022arXiv220806090L} found that in NGC 3627, the magnetic field inferred from synchrotron polarization agrees more with the magnetic fields traced by VGT-CO instead of the one traced by VGT-H$\alpha$, suggesting that synchrotron electron is well mixed with CO in star-forming regions. However, one should note that the acceleration and propagation of relativistic electrons responsible for synchrotron polarization in the centers of active galaxies is not a well-understood process. Therefore, the interpretation of synchrotron polarization angle is perpendicular to the POS magnetic field component can also be misleading.

\textbf{Prediction of magnetic field morphology towards NGC 4826:} 
We apply the VGT to obtain the first magnetic field map of NGC 4826, for which polarization measurements are not yet available. Fig.~\ref{ngc4826} shows the VGT-predicted magnetic field orientations using $\rm ^{12}CO$ emission line, which have a complex pattern in the central molecular gas reservoir. An hourglass-shaped magnetic field morphology is seen in the circumnuclear region. We find a transition from radial magnetic fields in the inner molecular gas disk with the radius $<0.75$ kpc to tangential magnetic fields in the outer region (see Fig.~\ref{AM_rad2} in the Supplementary). Coincidentally, streaming motions in CO were also observed in the inner disk \cite{Garc03}.

{\bf Dynamo activity:} The VGT-measured magnetic field morphology can help us understand the dynamo process in galaxies. For instance, preferentially tangential/spiral magnetic fields, which follow the differential rotation shear, are believed to be a product of the dynamo mechanism \cite{Beck96}. Such a globally spiral field is observed in M51, NGC 1068, NGC 1097's starburst ring, and NGC 3627's spiral arms (see the Supplementary Fig.~\ref{PM}). However, at the transition region from the inner bar to the outskirt, NGC 1097, NGC 1068, and NGC 3627 exhibit a preferentially radial magnetic field orientation. This change in magnetic field's morphology from tangential to radial suggests the gas's streaming motion becomes more important (see the Supplementary Fig.~\ref{PM}).
%In particular, the variations of PM and AM (VGT - HAWC+) show positive correlation (see the Supplementary Fig.~{\color{blue} 4}) suggesting the VGT and HAWC+ are tracing magnetic fields at different phases. %The change magnetic field's pitch angle (see the Supplementary Fig.~{\color{blue} 8}) presents a challenge for the existing dynamo theory \cite{2001ApJ...550..752V}.
In addition, dynamo studies in barred galaxies suggest dynamically more important magnetic fields and a closer alignment between magnetic and velocity fields compared to normal spiral galaxies \cite{Moss01}. The VGT-measured magnetic fields in CO show similar features, indicative of their dynamically important role in enhancing the molecular gas inflow in Seyfert galaxies. 

{\bf Implications of the observed alignment:} The agreement between the magnetic field orientations measured with the VGT and polarization reveals the coherence of magnetic fields across different gas phases, including the molecular phase traced by CO, denser phases traced by dust \cite{2015ARA&A..53..501A}, and the warm diffuse phase traced by synchrotron radiation. This alignment seen in Seyfert galaxies is also observed in the normal barred spiral galaxy Milky Way \cite{2020arXiv200715344A}. The observed coherence suggests that the magnetic fields threading the multiphase gas with an extended range of densities participate in the galactic dynamic evolution and undergo dynamo amplification \cite{2001ApJ...550..752V} and turbulent diffusion \cite{LV99}. It supports that molecular clouds are a part of the unified magnetic ecosystem in spiral galaxies and magnetic fields in diffuse and molecular phases have a coherent structure. This finding can have very important implications on many multi-scale processes, e.g., star formation \cite{Ching22}, cosmic ray propagation \cite{2011A&A...530A.109P}. 

{\bf Implications of the observed misalignment.}
Despite the alignment seen for the global magnetic structure especially in spiral arms, the nearby Seyfert galaxies feature the misalignment between the magnetic fields mapped with VGT-CO and polarization in regions with significant molecular gas inflows or, less likely outflows. This is particularly seen in the central part of M51, the bar-spiral interface of NGC 1068, and the nuclear of NGC 1097. In these regions with misalignment, accompanying the inflow, predominant radial magnetic fields in CO are also identified with the VGT. In NGC 3627, the misalignment marks the distortions in the molecular disk induced by a recent collision. As non-axisymmetric instabilities like bars and galaxy interactions are both drivers of gas inflow towards the nuclei \cite{Wada04}, our findings reveal the active role of magnetic fields in removing the angular momentum of molecular gas in the galactic disk and transporting it towards the nuclei of nearby Seyferts.

%\newpage
\section*{Methods}
\subsection*{Anisotropy of MHD turbulence}
The method of tracing magnetic fields through velocity gradients is rooted on the MHD turbulence theory (see the ref. \cite{GS95}, hereafter GS95) and turbulent reconnection theory (see the ref. \cite{LV99}, hereafter LV99). In what follows, we briefly explain those essentials. 

The prophetic study of GS95 proposed that the turbulent eddy is anisotropic, i.e., the eddy is elongating along magnetic field. Or in other word, the maximum velocity fluctuation appears in the direction perpendicular to the magnetic field. This anisotropy can be derived from the $''$critical balance$''$ condition: the cascading time ($k_\bot v_l$)$^{-1}$ equals the wave periods ($k_\parallel v_A$)$^{-1}$. Considering the velocity fluctuation is scale-dependent, for instance, the Kolmogorov-type turbulence $v_l\propto l^{1/3}$. Here $k_\parallel$ and $k_\bot$ are wavevectors parallel and perpendicular to the magnetic field, respectively. $v_l$ is turbulent velocity at scale $l$ and $v_A$ is Alfv\'{e}n speed. The corresponding GS95 anisotropy scaling is then: 
\begin{equation}
    k_\parallel\propto k_\bot^{2/3}
\end{equation}
which reveals the anisotropy increases as the scale of turbulent motions decreases. However, this derivation is drawn in Fourier space, in which the local spatial information is not available so that the anisotropy is measured with respect to the mean magnetic field, which builds up the global reference frame. In this frame, anisotropy of larger eddies dominates over small eddies
\cite{2000ApJ...539..273C}. Consequently, one can only observe a scale-independent anisotropy, which are dominated by the largest eddy \cite{2000ApJ...539..273C,2021ApJ...911...37H}. 

The scale-dependent anisotropic property of sub-Alf\'enic MHD turluence is later derived by LV99. LV99 explained that the scale-dependent anisotropy is only observable in the local reference frame, which is defined in real space with respect to the magnetic field passing through the eddy at scale $l$. In this frame, spatial information is available to define the local magnetic field. As local magnetic fields gives minimal resistance along the direction perpendicular to the magnetic field, it is easier to mix the magnetic field lines instead of bending it. Consequently, the turbulent cascading is channeled to the direction perpendicular to the magnetic field. Again, for instance, considering the Kolmogorov-type turbulence,
the motion of eddies perpendicular to the local magnetic field direction obeys the Kolmogorov law $v_{l,\bot}\propto l_\bot^{1/3}$. Here $l_\bot$ and $v_{l,\bot}$ are the perpendicular components of eddies' scale and velocity with respect to the {\it local} magnetic field, respectively. With the $''$critical balance$''$ in the local reference frame: $v_{l,\bot}l_\bot^{-1}\approx v_Al_\parallel^{-1}$, one can obtain the scale-dependent anisotropy scaling:
\begin{equation}
\label{eq.lv99}
 l_\parallel= L_{\rm inj}(\frac{l_\bot}{L_{\rm inj}})^{\frac{2}{3}}{M_A^{-4/3}}, { M_A\le 1}
\end{equation}
where $l_\|$ is the parallel component of eddies' scale. $L_{\rm inj}$ is the turbulence injection scale and $M_A$ is Alfv\'en Mach number. This scale-dependent anisotropy in the local reference frame was numerically demonstrated
\cite{2000ApJ...539..273C,2001ApJ...554.1175M} and in-situ observations in solar wind \cite{2016ApJ...816...15W,2021ApJ...915L...8D,2020FrASS...7...83M}. 

Super-Alfv\'enic MHD turbulence (i.e., $M_{\rm A}>1$) is typically isotropic. The Super-Alfv\'enic motions at the injection scale are hydrodynamic due to the relatively weak backreaction of the magnetic field. However, the kinetic energy of turbulent motions follows the nearly isotropic turbulent cascade. The importance of magnetic backreaction gets stronger at smaller scales. Eventually, at the scale $l_{\rm A} = L_{\rm inj}M_{\rm A}^{-3}$, the turbulent velocity becomes equal to the Alfv\'en velocity so that the turbulence is anisotropic \cite{2006ApJ...645L..25L}. 

In particular, the corresponding anisotropy scaling for velocity fluctuation and gradient's amplitude of velocity fluctuation are  \cite{LV99}:
\begin{equation}
\label{eq.vg}
\begin{aligned}
 v_{l,\bot}&= v_{\rm inj}(\frac{l_\bot}{L_{\rm inj}})^{\frac{1}{3}}{M_A^{1/3}}\\
      \nabla v_l&\propto\frac{v_{l,\bot}}{l_\bot}\simeq \frac{v_{\rm inj}}{L_{\rm inj}}(\frac{l_{\perp}}{L_{\rm inj}})^{-\frac{2}{3}}M_A^{\frac{1}{3}}
\end{aligned}
\end{equation}
where $v_{\rm inj}$ is the injection velocity. $\nabla v_l$ 's direction points to the maximum changes of velocity fluctuation, i.e., perpendicular to the local magnetic field.

A vital property of turbulence's gradient induced by turbulence is the increase of gradient amplitude with the decrease of the scale \cite{CMZ}, which is not usually true for large-scale gradient of non-turbulent nature, for instance, the galactic different rotation. When dealing with Seyfert galaxies, it is therefore necessary that the observation resolves the turbulence's injection scale $\sim100$~pc so that the turbulent gradient is dominated, although this does not exclude that in some observations gradients cannot arise due to other reasons. 

\subsection*{The VGT}
The VGT is the main analysis tool in the work. As introduced above, it is theoretically rooted in the advanced magnetohydrodynamic (MHD) turbulence theory \cite{GS95} and fast turbulent reconnection theory \cite{LV99}. Here we follow the VGT recipe proposed in \cite{CMZ} employing thin velocity channel maps \textbf{Ch(x,y)} to extract velocity information in Position-Position-Velocity (PPV) cubes via the velocity caustics effect. The concept of velocity caustics effect was proposed by the ref.~\cite{LP00} to signify the effect of density structure distortion due to turbulent and shear velocities along the LOS. Since the density structure with different velocities is sampled into different velocity channels, the density structure is significantly modified. The statistics of the intensity fluctuations in PPV and their relations to the underlying statistics of turbulent velocity and density are formulated in the ref.~\cite{LP00}. The study showed that the velocity fluctuations are most prominent in thin channel maps \cite{LP00,2021ApJ...915...67H}. In particular, the thin channel and the thick channel can be distinguished by:
\begin{equation}
\label{eq1}
\begin{aligned}
    \Delta v&<\sqrt{\delta (v^2)}, \thickspace\mbox{thin channel},\\
    \Delta v&\ge\sqrt{\delta (v^2)},\thickspace\mbox{thick channel},
\end{aligned}
\end{equation}
where $\Delta v$ is the velocity channel width, $\sqrt{\delta (v^2)}$ is the velocity dispersion. The choice of channel thickness is intended to increase the relative weight of velocity-induced fluctuations compared to the density-induced ones \cite{LP00}. This is important as, due to the properties of MHD turbulence, velocity fluctuations are better aligned with magnetic field compared to their density counterparts. The validity of velocity caustics effect in the multiple-phase medium of neutral hydrogen H I was questioned by the ref.~\cite{2019ApJ...874..171C}. However, their arguments were exposed by the ref.~\cite{2019arXiv190403173Y} with the analysis of observational data in the ref.~\cite{2021ApJ...910..161Y}. It demonstrates the importance of velocity caustics in the multi-phase galactic H I.

To calculate velocity gradients, each thin channel map is convolved with 3 $\times$ 3 Sobel kernels $G_x=\begin{pmatrix} 
	-1 & 0 & +1 \\
	-2 & 0 & +2 \\
	-1 & 0 & +1
	\end{pmatrix}$ and $G_y=\begin{pmatrix} 
	-1 & -2 & -1 \\
	0 & 0 & 0 \\
	+1 & +2 & +1
	\end{pmatrix}$ to calculate pixelized gradient map $\psi_{\rm g}^i(x,y)$:
\begin{equation}
\label{eq:conv}
\begin{aligned}
\bigtriangledown_x {\rm Ch}_i(x,y)&=G_x * {\rm Ch}_i(x,y)\\ 
\bigtriangledown_y {\rm Ch}_i(x,y)&=G_y * {\rm Ch}_i(x,y)\\
\psi_{\rm g}^i(x,y)&=\tan^{-1}\left(\frac{\bigtriangledown_y {\rm Ch}_i(x,y)}{\bigtriangledown_x {\rm Ch}_i(x,y)}\right),
\end{aligned}
\end{equation}
where $\bigtriangledown_x {\rm Ch}_i(x,y)$ and $\bigtriangledown_y {\rm Ch}_i(x,y)$ are the $x$ and $y$ components of gradient, respectively, and $*$ denotes the convolution. To suppresses the effect noise in the spectroscopic data, we mask out the raw gradient whose corresponding intensity is less than three times of the RMS noise level.

As the orthogonal relative orientation between velocity gradients and the magnetic field appears only when the sampling is statistically sufficient, each raw gradient map $\psi_{\rm g}^i(x,y)$ is further processed by the sub-block averaging method \cite{YL17a}. The sub-block averaging method takes all gradients orientation within a sub-block of interest and then plots the corresponding histogram. A Gaussian fitting is then applied to the histogram. The Gaussian distribution's peak value gives the statistically most probable gradient's orientation within that sub-block.

By repeating the gradient's calculation and the sub-block averaging method for each thin velocity channel, we obtain totally $n_v$ processed gradient maps $\psi_{\rm g,s}^i(x,y)$ with $i=1,2,...,n_v$. In analogy to the Stokes parameters of polarization, the pseudo $Q_{\rm g}$ and $U_{\rm g}$ of gradient-inferred magnetic fields are defined as:
\begin{equation}
\label{eq.qu}
\begin{aligned}
& Q_{\rm g}(x,y)=\sum_{i=1}^{n_v} {\rm Ch}_i(x,y)\cos(2\psi_{\rm g,s}^i(x,y)),\\
& U_{\rm g}(x,y)=\sum_{i=1}^{n_v} {\rm Ch}_i(x,y)\sin(2\psi_{\rm g,s}^i(x,y)),\\
& \psi_{\rm g}(x,y)=\frac{1}{2}\tan^{-1}(\frac{U_{\rm g}}{Q_{\rm g}}),
\end{aligned}
\end{equation}
The pseudo polarization angle $\psi_{\rm g}$ is then defined correspondingly. Similar to the Planck polarization, $\psi_{\rm B}=\psi_{\rm g}+\pi/2$ gives the POS magnetic field orientation. In this work, we do not make a selection of velocity range but use all emissions from the galaxies. The use of velocity caustic effect and pseudo-Stokes parameters distinguish the VGT from typical structure or edge detection algorithm via gradient used in imaging processing. Due to the velocity caustic effect, the gradient of the emission in thin channels does not necessarily detect real density structures in the galaxy. By integrating the pseudo-Stokes parameters along the LOS, the VGT reflects the emission's dynamic information rather than the emission's total intensity contours.

\subsection*{Observational data}
\subsubsection*{Emission lines}
The $\rm ^{12}CO$ (J = 1-0), $\rm ^{13}CO$ (J = 1-0), and $\rm ^{12}CO$ (J = 2-1) emission lines used in this work come from the PdBI Arcsecond Whirlpool Survey (PAWS)\cite{2013ApJ...779...46H}, ALMA NGC 1068 project \cite{2017PASJ...69...18T}, and the Physics at High Angular resolution in Nearby Galaxies-Atacama Large Millimeter/submillimeter Array (PHANGS–ALMA) survey \cite{2021arXiv210407739L}, respectively. Summary of data set is presented in the Supplementary Tab.~{\color{blue} 1}.

\textbf{M51:} The PAWS maps the $\rm ^{12}CO$ (J = 1-0) emission from the central $\sim9$ kpc of a massive spiral galaxy M51. The effective angular resolution of the $\rm ^{12}CO$ data cube is $1.16'' \times 0.97''$, corresponding to a spatial resolution of $\sim37$ pc. The mean RMS of the noise is $\sim0.4$ K in a 5.0 km s$^{-1}$ channel.

\textbf{NGC 1068:} The central $\sim1'$ diameter region of NGC 1068 was observed with $\rm ^{13}CO$ (J = 1-0) using the Band 3 receiver on ALMA. The effective beam sizes is $1.4''\times1.4''$, giving a  spatial resolution of $\sim98$ pc. The resultant mean RMS noise level in the channel maps is $\sim0.64$ mJy beam$^{-1}$ at a velocity resolution of 1.5 km s$^{-1}$.

\textbf{NGC 1097, NGC 3627, and NGC 4826:} PHANGS-ALMA survey provides $\rm ^{12}CO$ (J = 1-0) emission lines emission at $\sim1''$ ($\sim$100pc) spatial resolution and 2.5 km s$^{-1}$ velocity resolution for the nearby galaxies NGC 1097, NGC 3627, and NGC 4826. The survey achieves a high signal-to-noise ratio with an RMS brightness temperature noise level of $\sim0.30\pm0.13$ K km s$^{-1}$.

\subsubsection*{Polarization measurement}
We use the HAWC+ polarization measurement obtained from HAWC+ archival database \cite{2018JAI.....740008H,M51,2020ApJ...888...66L}. The magnetic field orientation is defined as $\phi_B=\phi+\pi/2$, inferred from the polarization angle $\phi$ along with the polarization fraction $p$, specifically:
\begin{equation}
\begin{aligned}
\phi&=\frac{1}{2}\arctan(U,Q)\\
p&=\sqrt{Q^2+U^2}/I
        \end{aligned}
        \end{equation}
where $I$, $Q$, and $U$ refer to the intensity of dust emission and Stokes parameters, respectively. We use the band D measurement (154 $\mu$m, FWHM $\approx13.6''$) for galaxies M51 and NGC 3627, while band C measurement (89 $\mu$m, FWHM $\approx7.8''$) for  and NGC 1068/NGC 1097.  We  also adopt the VLA radio observation (combined with Effelsberg telescope) at wavelength 6.2 cm (FWHM $\approx8.0''$) for M51\cite{2011MNRAS.412.2396F} and 3.5 cm (FWHM $\approx3.0''$) for NGC 1097\cite{2005A&A...444..739B}, respectively. We consider only pixels with $p/\sigma_p > 2$, where $p$ is the polarization fraction, and $\sigma_p$ is its uncertainty. The polarization data is not smoothed further but re-grid to match that of the VGT-measured magnetic field orientation. 

\section*{Data availability}
The data that support the plots within this paper and other findings of this study are available from the corresponding authors and other co-authors upon reasonable request.

\section*{Code availability}
The codes support the plots within this paper are available from the corresponding authors upon reasonable request.

\bibliography{sample}

\section*{Acknowledgements}
Y.H. acknowledges the support of the NASA TCAN 144AAG1967. A.L. acknowledges the support of the NASA ATP AAH7546. S.X.~acknowledges the support for this work provided by NASA through the NASA Hubble Fellowship grant \# HST-HF2-51473.001-A awarded by the Space Telescope Science Institute, which is operated by the Association of Universities for Research in Astronomy, Incorporated, under NASA contract NAS5-26555. We thank James Stone for helpful discussions. This work made use of PAWS, 'The PdBI Arcsecond Whirlpool Survey'. Based M51, NGC 1068, and NGC 3627 on observations made with the NASA/DLR Stratospheric Observatory for Infrared Astronomy (SOFIA). SOFIA is jointly operated by the Universities Space Research Association, Inc. (USRA), under NASA contract NNA17BF53C, and the Deutsches SOFIA Institut (DSI) under DLR contract 50 OK 0901 to the University of Stuttgart. This work made use of PHANGS–ALMA, ’the Physics at High Angular resolution in Nearby Galaxies-Atacama Large Millimeter/submillimeter Array survey’. ALMA is a partnership of ESO (representing its member states), NSF (USA) and NINS (Japan), together with NRC (Canada), MOST and ASIAA (China, Taiwan), and KASI (Republic of Korea), in cooperation with the Republic of Chile. The Joint ALMA Observatory is operated by ESO, AUI/NRAO and NAOJ. The National Radio Astronomy Observatory is a facility of the National Science Foundation operated under cooperative agreement by Associated Universities, Inc. This paper makes use of the following ALMA data: ADS/JAO.ALMA\#2012.1.00650.S, 2013.1.00803.S, 2013.1.01161.S, 2015.1.00121.S, 2015.1.00782.S, 2015.1.00925.S, 2015.1.00956.S, 2016.1.00386.S, 2017.1.00392.S, 2017.1.00766.S, 2017.1.00886.L, 2018.1.00484.S, 2018.1.01321.S,\\ 2018.1.01651.S, 2018.A.00062.S, 2019.1.01235.S, 2019.2.00129.S.

\section*{Author contributions}
All authors discussed the results, commented on the manuscript and contributed to the writing of the manuscript. Y.H. and A.L. conceived the project, Y.H. performed calculations, while Y.H. and A.L. analysed the results and wrote the original manuscript. R.B. provided the VLA radio data of M51/NGC 1097 and suggestions on interpolating the magnetic fields in extragalaxies. S.X. provided explanations of the observed differences in magnetic field directions obtained with the VGT and the polarization. 

\section*{Competing interests}
The authors declare no competing interests. 
%\begin{figure}[ht]
%\centering
%\includegraphics[width=0.99\linewidth]{optical.png}
%\caption{The composite images of the five galaxies from HST and Chandra archives. Credit: ESA/Hubble \& NASA, J. Lee and the PHANGS-HST Team, NASA/CXC/Ohio State Univ./C.Grier et al., NASA/STScI, ESO/WFI, NASA/JPL-Caltech, NASA/CXC/MIT/C.Canizares, D.Evans et al, NASA/STScI, NSF/NRAO/VLA.
%}
%\label{optical}
%\end{figure}
\newpage
%\section*{Box 1 (Optional)}
%This is a Box, which can contain a figure, and which should have no more than 300 words of text.

%\begin{figure}[ht]
%\centering
%\includegraphics[width=\linewidth]{fig}
%\caption{The figure caption should start with a title explaining the figure. Figures should be self-consistent so please redefine all acronyms and define all symbols. Example: GHZ, Greenberger–Horne–Zeilinger, OAM, orbital angular momentum. Please provide credit lines for panels reproduced from the literature. Example: Panels a and b are reproduced from Ref. \cite{TR}.}
%\label{fig}
%\end{figure}

\section*{Supplementary information}
\section{Distribution of AM}
\begin{figure}[ht]
\centering
\includegraphics[width=1.0\linewidth]{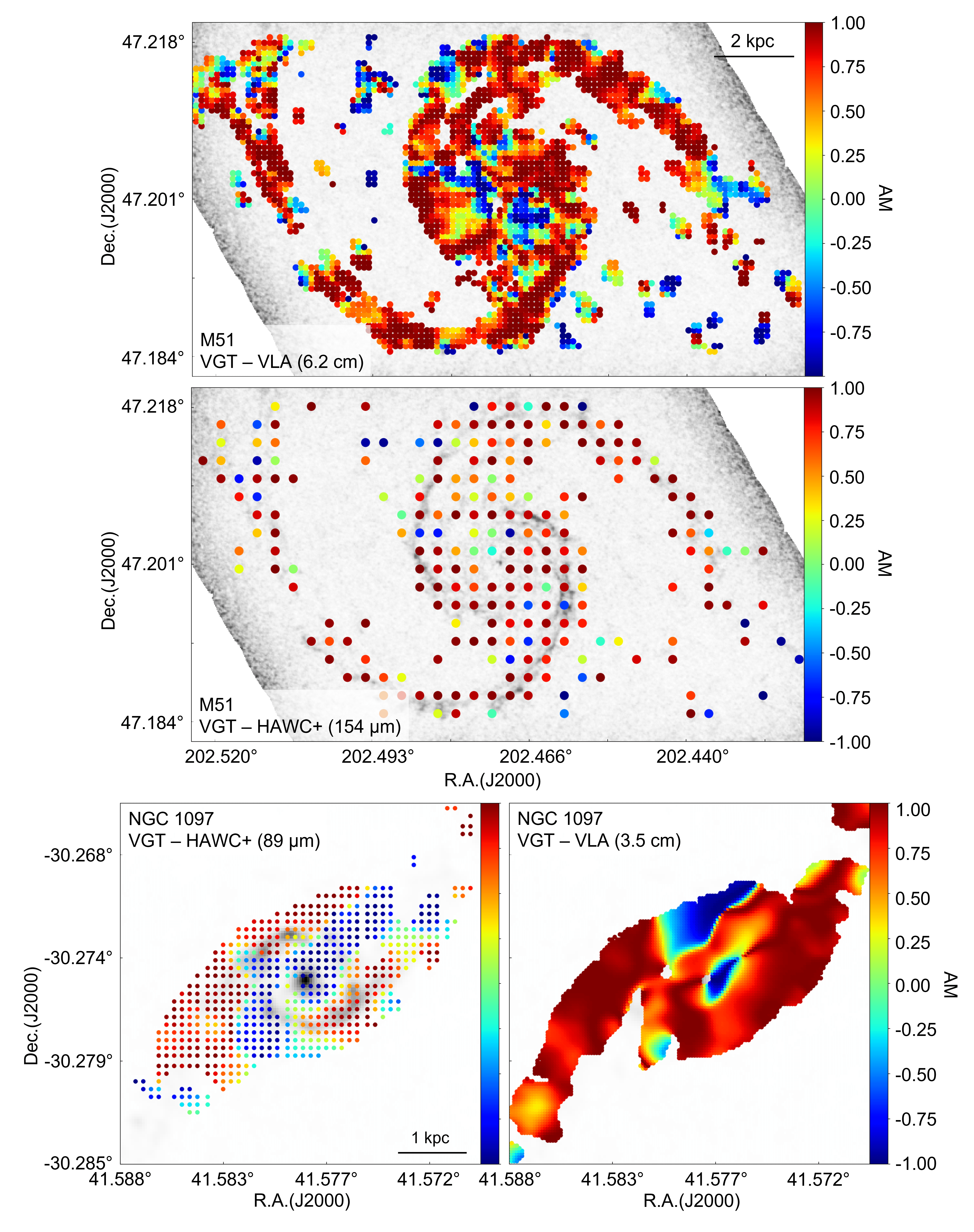}
\caption{Distribution of AM (between the VGT and polarization) towards M51 (top; dust polarization), NGC 197 (bottom left: dust polarization; bottom right: synchrotron polarization) galaxies.
 All plots share the same colorbar and overlaid on CO emission intensity maps.}
\label{AM_dis1}
\end{figure}

\begin{figure}[ht]
\centering
\includegraphics[width=1.0\linewidth]{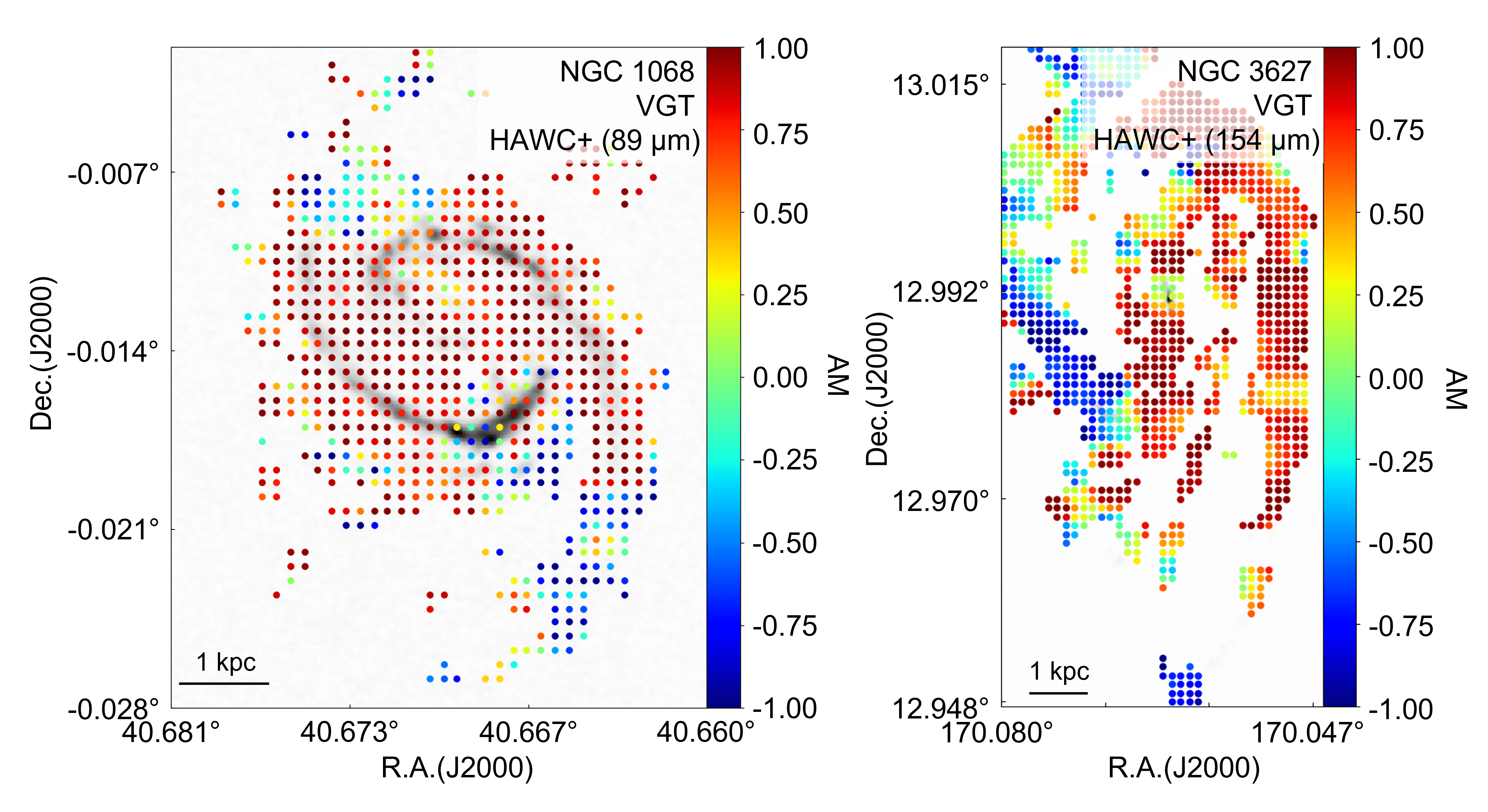}
\caption{Distribution of AM (between the VGT and polarization) towards NGC 1068 (left; dust polarization) and NGC 1097 (right; synchrotron polarization) galaxies. The background images and contours are CO emission intensity maps.}
\label{AM_dis2}
\end{figure}

\begin{figure}[ht]
\centering
\includegraphics[width=1.0\linewidth]{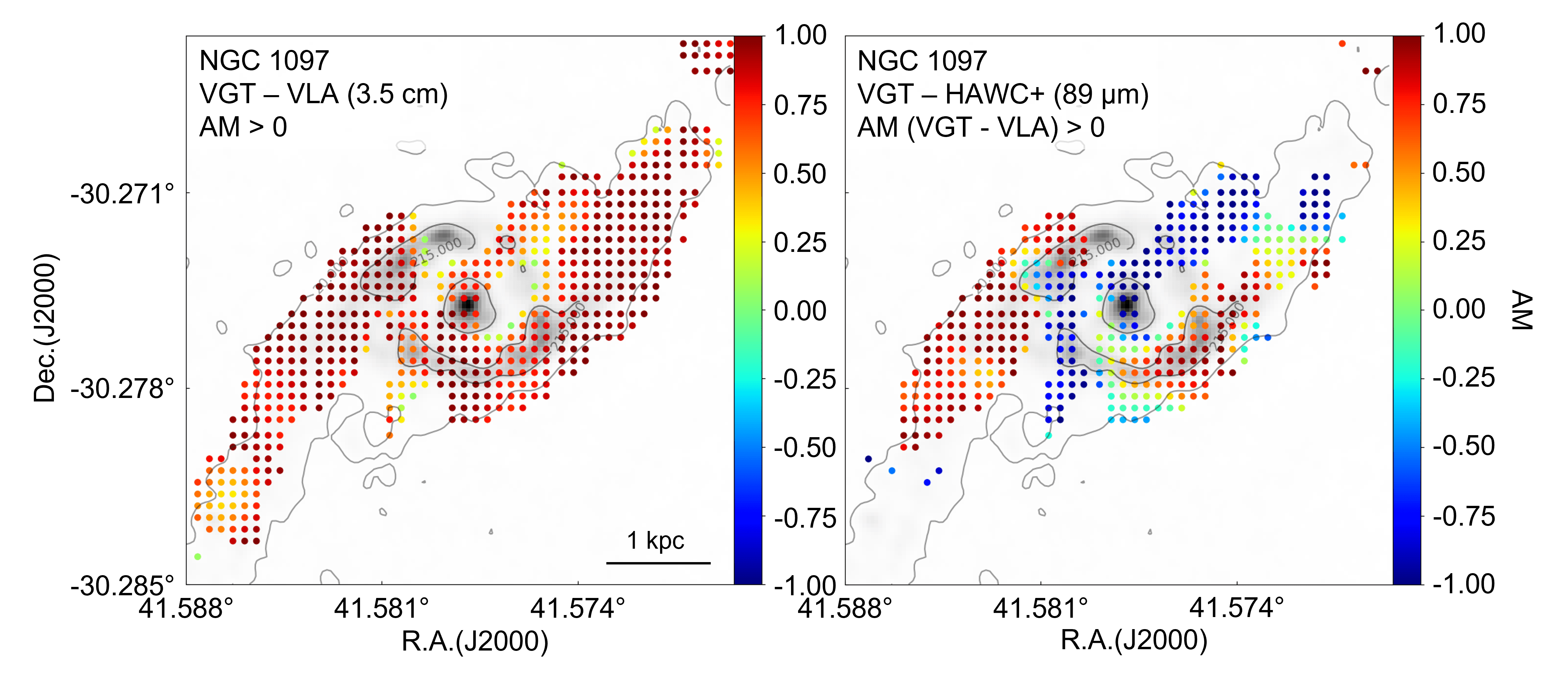}
\caption{Distribution of AM (between the VGT and polarization) towards the zoom-in central part of NGC 1097. For both AM maps of VGT - VLA (right) and VGT - HAWC+ (left), we blanked out negative AM (VGT - VLA) values. The background image and contours are CO emission intensity maps.}
\label{AM_zoom}
\end{figure}

\begin{figure}[t]
\centering
\includegraphics[width=1.0\linewidth]{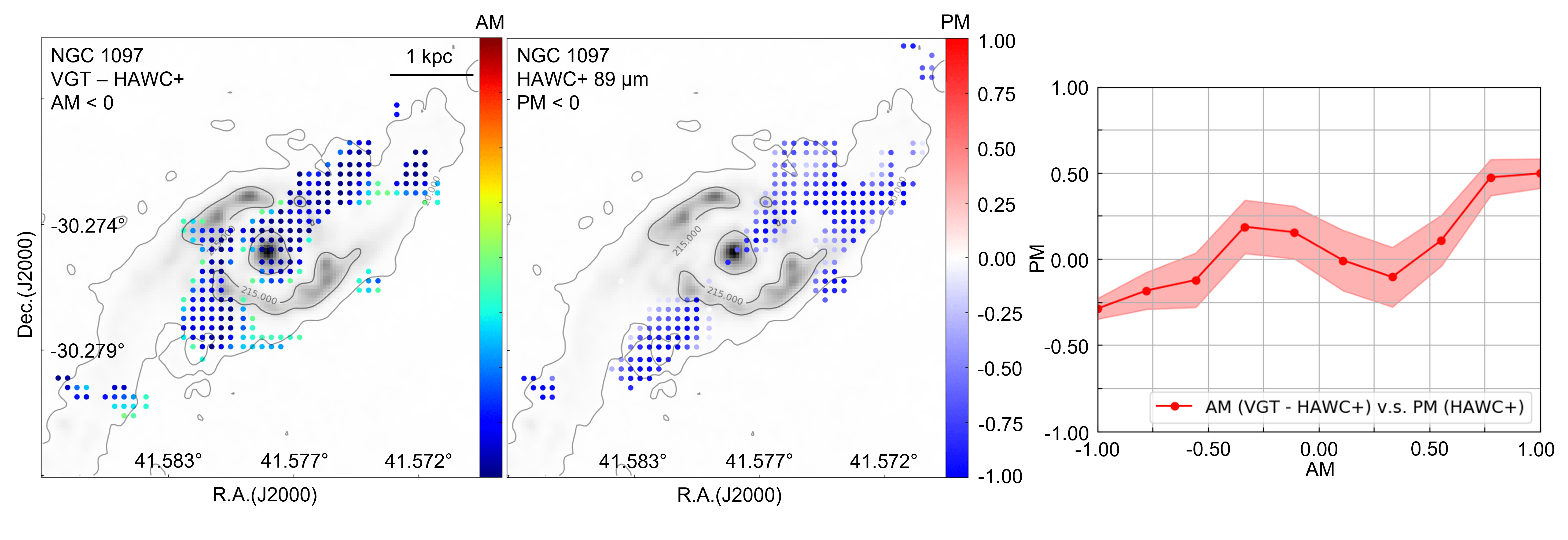}
\caption{\textbf{Left:} distribution of AM (between the VGT and HAWC+ polarization) towards the starburst ring of NGC 1097. Positive AM values (i.e., agreement of the VGT and HAWC+) are blanked out. The colorbar has the same range [-1, 1] as the middle panel. \textbf{Middle: } distribution of HAWC+'s PM towards the same region. Positive PM values (i.e., tangential field) are blanked out. \textbf{Right:} the correlation of AM and PM without blanking out any values. The PM is averaged over uniformly spaced AM bins. The shadow area gives the standard deviation of the PM.
}
\label{ampm}
\end{figure}

\begin{figure}[ht]
\centering
\includegraphics[width=0.95\linewidth]{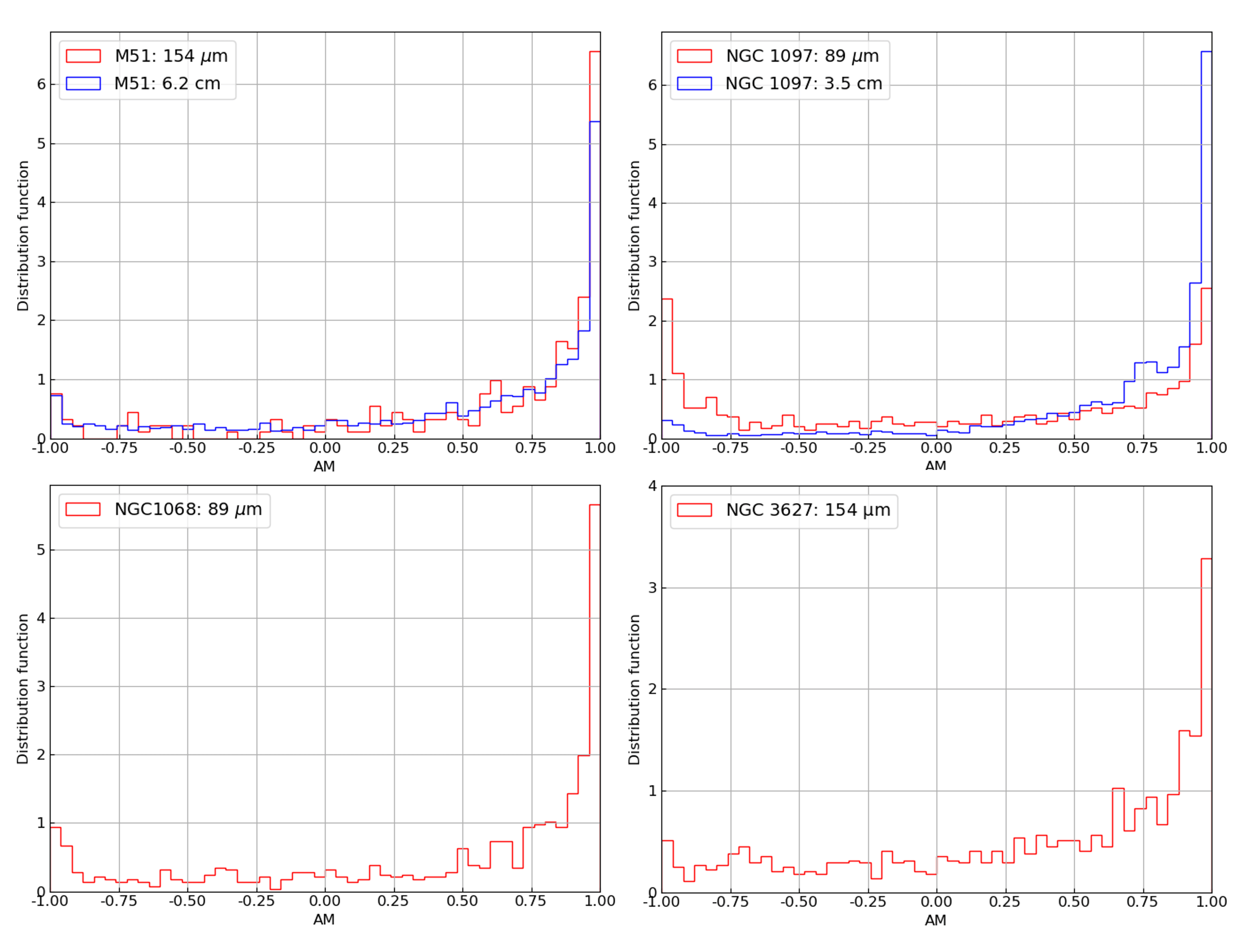}
\caption{Histograms of AM (between the VGT and polarization) towards M51 (top left), NGC 1097 (top right dust), NGC 0168 (bottom left), and NGC 3627 (bottom right) galaxies. Red line indicates dust polarization, while blue line means synchrotron polarization.}
\label{AM_hist}
\end{figure}

We present the distribution maps of AM value (${\rm AM}=2(\cos^{2} \theta_{r}-\frac{1}{2})$, where $\theta_r$ is the relative angle between the two magnetic field vectors) in Figs.~\ref{AM_dis1} and ~\ref{AM_dis2}. AM = 1 means two vectors are along the same direction, while AM = -1 suggests two vectors are perpendicular to each. In general, the magnetic fields traced by the VGT are globally compatible with the polarization measurements (i.e., AM$\sim1$). Although discrepancy exists, a good agreement suggests turbulence's role is important. Turbulence's intrinsic properties may contribute to the discrepancy, because super-Alfv\'enic turbulence is isotropic.  It is possible that the telescope does not resolve the scale $l_{\rm A}$ in some super-Alfv\'enic regions, so we observe a disagreement of the VGT and polarization. In addition, dust is well mixed with all the interstellar phases, including both molecular and atomic gas \cite{2015ARA&A..53..501A}. Dust size distribution, alignment, as well as the magnetic field direction, can change in different phases. For instance, shocks and radiative torque disruption change the size distribution of dust. This affects dust alignment \cite{2021ApJ...908...12L}. Therefore, the polarization from dust depends on the intensity of radiation in the wavelength comparable with the size of the grains. The alignment of dust has never been studied in the environment of active galaxies. Therefore, it may not be surprising that we see the differences between the dust polarization sampling aligned dust in very different conditions along the LOS, and the VGT-CO. 

As both the VGT and HAWC+ sample the magnetic field in the cold-gas phase, their agreement is expected as observed in M51, NGC 1068, and NGC 3627. However, strikingly in NGC 1097, the VGT significantly differs from HAWC+ dust polarization but agrees with VLA synchrotron polarization, which measures the ionized gas phase. To investigate this unexpected discrepancy, we blank out the pixels in which the AM of VGT - VLA alignment is negative. As shown in Fig.~\ref{AM_zoom}, the discrepancy mostly appears in the central disk region, including the upper and lower parts of the starburst ring.

Fig.~\ref{AM_hist} shows the histograms of AM (between the VGT and polarization) towards the four galaxies. While the distribution spans from -1 to 1, the majority concentrates on the range of 0.75 - 1.0 still. It suggests a globally compatible agreement between the VGT and polarization.  

\begin{figure}[ht]
\centering
\includegraphics[width=1.00\linewidth]{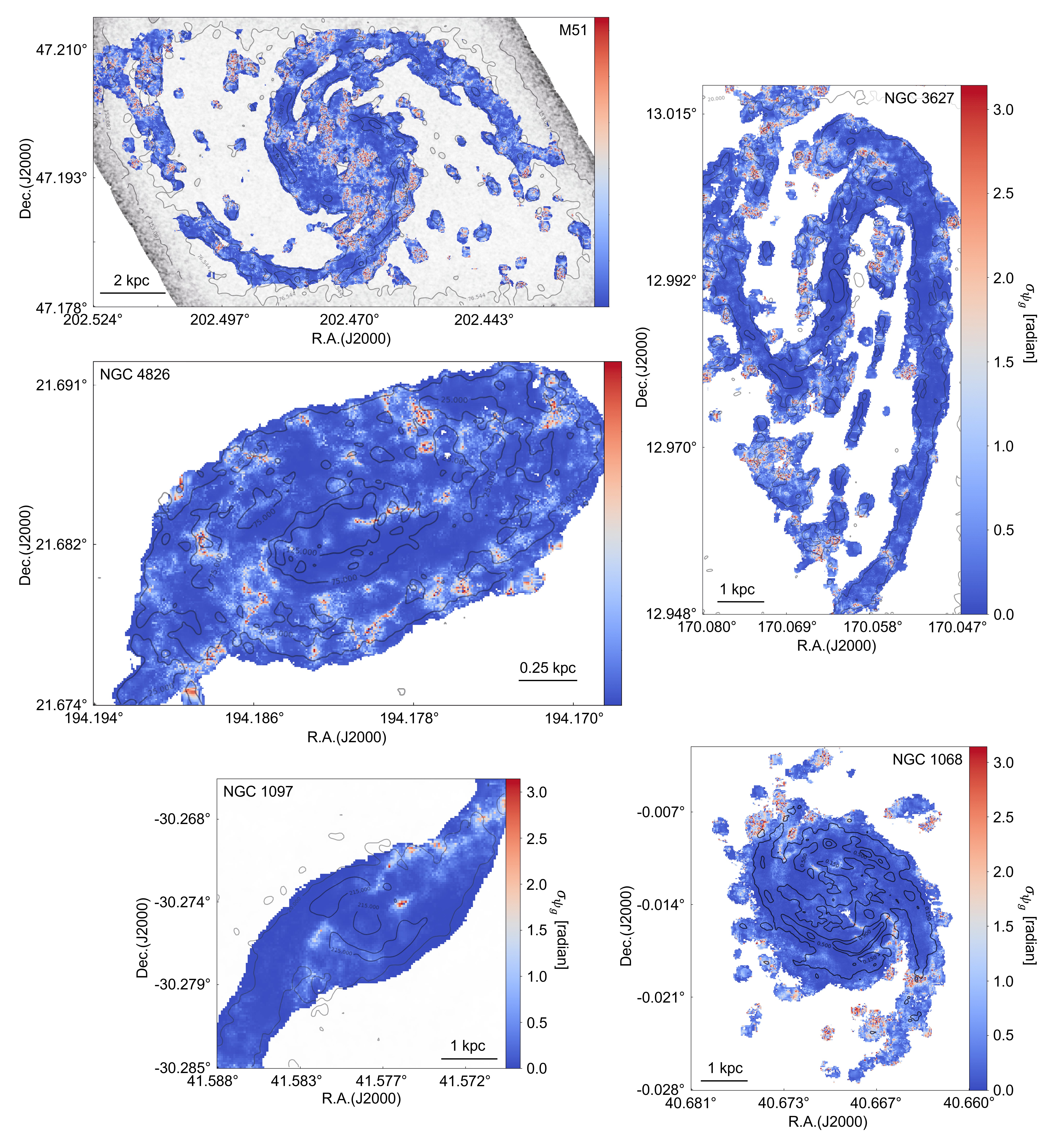}
\caption{Uncertainty maps for the magnetic field measured by the VGT for the five galaxies M51, NGC 1068, NGC 1097, NGC 3627, and NGC 4826. All plots share the same colorbar.  The background image and contours are CO emission intensity maps.}
\label{error1}
\end{figure}

\section{Uncertainty of the magnetic field direction measured by the VGT}
The two significant uncertainties of the magnetic field can come from the systematic error in the observational map and the VGT algorithm. For the latter, the VGT takes a subregion and fits a corresponding Gaussian histogram of the gradient’s orientation. It then outputs the angle of orientation corresponding to the Gaussian fitting peak value of the histogram. The uncertainty therefore can be considered as the error $\sigma_{\psi_{\rm gs}}(x,y,v)$ from the Gaussian fitting algorithm within the 95\% confidence level. 

Considering the noise $\sigma_n(x,y,v)$ in velocity channel Ch$(x,y,v)$ and error propagation, the uncertainties $\sigma_{Q}(x,y)$ and $\sigma_{U}(x,y)$ of the Pseudo Stokes parameters $Q_{\rm g}(x,y)$ and $U_{\rm g}(x,y)$ can be obtained from:
\begin{equation}
    \begin{aligned}
    \sigma_{\cos}(x,y,v)&=|2\sin(2\psi_{\rm gs}(x,y,v))\sigma_{\psi_{\rm gs}}(x,y,v)|\\
    \sigma_{\sin}(x,y,v)&=|2\cos(2\psi_{\rm gs}(x,y,v))\sigma_{\psi_{\rm gs}}(x,y,v)|\\
    \sigma_{q}(x,y,v)&=|\rm{Ch}\cdot\cos(2\psi_{\rm gs})|\sqrt{(\sigma_n/\rm{Ch})^2+(\sigma_{\cos}/\cos(2\psi_{\rm g}))^2}\\
    \sigma_{u}(x,y,v)&=|\rm{Ch}\cdot\sin(2\psi_{\rm gs})|\sqrt{(\sigma_n/\rm{Ch})^2+(\sigma_{\sin}/\sin(2\psi_{\rm g}))^2}\\
    \sigma_{Q}(x,y)&=\sqrt{\sum_{v}\sigma_{q}(x,y,v)^2}\\
    \sigma_{U}(x,y)&=\sqrt{\sum_{v}\sigma_{u}(x,y,v)^2}\\
    \sigma_{\psi_{\rm g}}(x,y)&=\frac{|U_{\rm g}/Q_{\rm g}|\sqrt{(\sigma_{Q}/Q_{\rm g})^2+(\sigma_{U}/U_{\rm g})^2}}{2[1+(U_{\rm g}/Q_{\rm g})^2]}
    \end{aligned}
\end{equation}
where $\sigma_{\psi_{\rm g}}(x,y)$ gives the angular uncertainty of the resulting magnetic field direction. The uncertainty maps of the galaxies are presented in Figs.~\ref{error1}. The median value is listed in Tab.~\ref{tab.2}. 

\begin{table}[ht]
\centering
\begin{tabular}{|c|c|c|c|c|c|c|}
\hline
Galaxy & Distance & Resolution & Emission line & Data source & $\Delta v$ & Polarization\\
\hline
M51 & 7.6 Mpc & 37.0 pc & $\rm ^{12}CO$(J = 1-0) & PAWS\cite{2013ApJ...779...46H}  & 5.0 km s$^{-1}$ & HAWC+ 154 ${\rm \mu m}$\cite{M51} \& VLA 6.2 cm \cite{2011MNRAS.412.2396F}\\
NGC 1068 & 14.4 Mpc & 98.0 pc & $\rm ^{13}CO$(J = 1-0) & ALMA \cite{2017PASJ...69...18T} & 1.5 km s$^{-1}$ & HAWC+ 89 $\rm \mu m$ \cite{2020ApJ...888...66L}\\
NGC 1097 & 17.0 Mpc & 82.0 pc & $\rm ^{12}CO$(J = 2-1) &  PHANGS–ALMA\cite{2021arXiv210407739L,2021ApJS..255...19L}  & 2.5 km s$^{-1}$ & HAWC+ 89 $\rm \mu m$ \cite{2021arXiv210709063L} \& VLA 3.5 cm \cite{2005A&A...444..739B}\\
NGC 3627 & 9.6 Mpc & 47.0 pc & $\rm ^{12}CO$(J = 2-1) & PHANGS–ALMA\cite{2021arXiv210407739L,2021ApJS..255...19L} & 2.5 km s$^{-1}$ & HAWC+ 154 $\rm \mu m$\\
NGC 4826 & 5.3 Mpc & 26.0 pc & $\rm ^{12}CO$(J = 2-1) & PHANGS–ALMA\cite{2021arXiv210407739L,2021ApJS..255...19L} & 2.5 km s$^{-1}$ & -\\
\hline
\end{tabular}
\caption{\label{tab}Summary of data sets used in this work. $\Delta v$ represents the velocity resolution of emission lines.}
\end{table}

\begin{table}[ht]
\centering
\begin{tabular}{|c|c|c|c|c|c|c|}
\hline
Galaxy & R.A.(J2000) & Dec. (J2000) & $i$ & P.A. & $\sigma_{\psi_{g}}$ & $\langle\theta_p\rangle$ (VGT, HAWC+, VLA) \\
\hline
M51\cite{2013ApJ...762L..27H}  & 202.469$^\circ$ & 47.195$^\circ$ & 20.3$^\circ$ & 12.0$^\circ$ & 14.99$^\circ$ & 24.98$^\circ\pm0.12^\circ$, 26.54$^\circ\pm1.34^\circ$, 28.32$^\circ\pm0.38^\circ$ \\
NGC 1068\cite{2019A&A...624A..80N,2020ApJ...888...66L} & 40.670$^\circ$ & -0.013$^\circ$ & 48.1$^\circ$ & 52.0$^\circ$ & 10.50$^\circ$&16.26$^\circ\pm0.14^\circ$, 31.28$^\circ\pm1.11^\circ$, -\\
NGC 1097\cite{2005A&A...444..739B} & 41.579$^\circ$ & -30.275$^\circ$ & 45.0$^\circ$ & -45.0$^\circ$ & 12.37$^\circ$& -30.60$^\circ\pm0.35^\circ$, -8.80$^\circ\pm1.98^\circ$, -38.47$^\circ\pm0.35^\circ$\\
NGC 3627\cite{2017A&A...597A..85B} & 170.063$^\circ$ & 12.991$^\circ$ & 65.0$^\circ$ & 170.0$^\circ$ & 11.30$^\circ$ & 11.48$^\circ\pm0.12^\circ$, -7.77$^\circ\pm1.31^\circ$, -\\
NGC 4826\cite{2003A&A...407..485G} & 194.182$^\circ$ & 21.683$^\circ$ & 60.0$^\circ$ & 112.0$^\circ$ & 8.92$^\circ$& 14.11$^\circ\pm0.24^\circ$, -, -\\
\hline
\end{tabular}
\caption{\label{tab.2}Summary of galaxies' parameters used in this work. The inclination ($i$) of the galaxy disk is measured with respect to the POS and the position angle (P.A.) of the major axis of the projected galaxy disk in the sky plane is measured in IAU convention. $\sigma_{\psi_{g}}$ represents the median uncertainty of the VGT measurement and $\langle\theta_p\rangle$ is the mean pitch angle .}
\end{table}

\begin{figure}[b]
\centering
\includegraphics[width=0.99\linewidth]{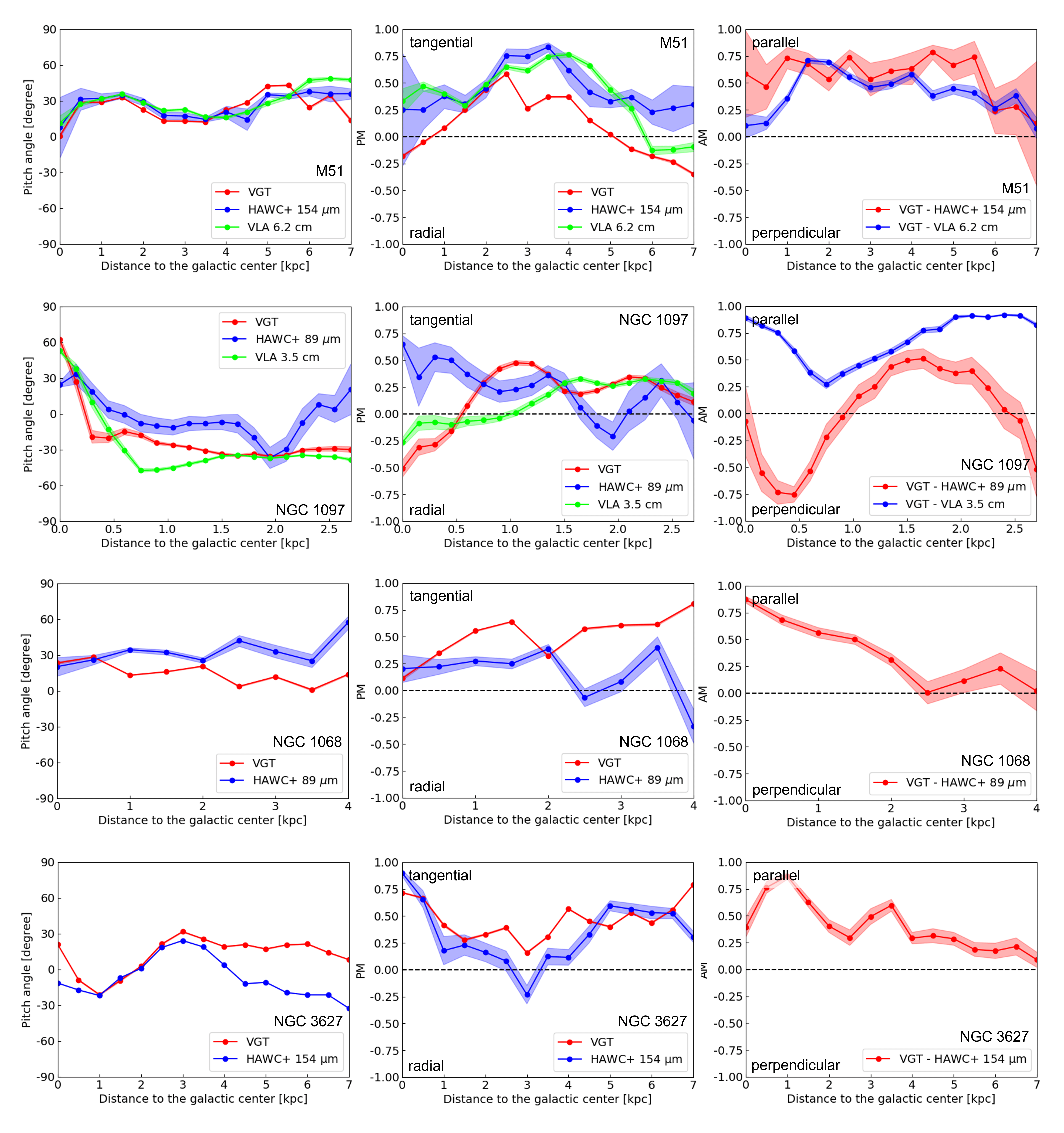}
\caption{\textbf{Left:} pitch angle as a function of the distance (x-axis) to the galactic center towards the M51, NGC 1097, NGC 1068, and NGC 3627 galaxies. \textbf{Middle:} PM as a function of the distance to the galactic center. PM > 0 indicates a preferentially tangential field and PM < 0 suggests a preferentially radial field. \textbf{Right:} AM as a function of the distance to the galactic center. AM = 1 means two measurement are identical, while AM = -1 suggests orthogonal relative angle.
}
\label{AM_rad}
\end{figure}

\begin{figure}[b]
\centering
\includegraphics[width=0.95\linewidth]{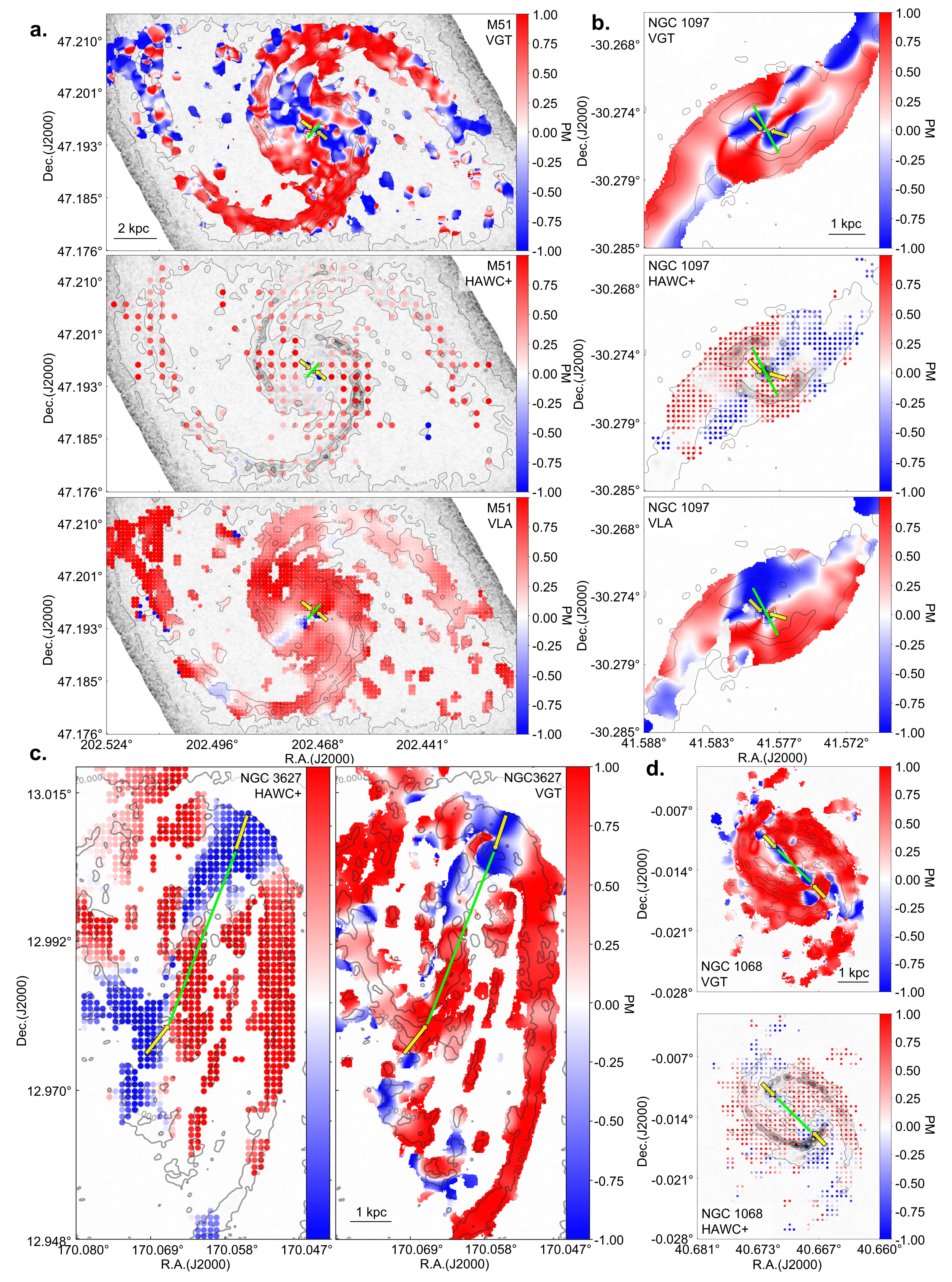}
\caption{The distribution of PM in M51 (panel a), NGC 1097 (panel b), NGC 3627 (panel c), and NGC 1068 (panel d) galaxies. PM > 0 indicates a preferentially tangential field and PM < 0 suggests a preferentially radial field. Approximated locations of the inner-bar are labelled by green segment and yellow arrows indicate the direction of inflows. The background images and contours are CO emission intensity maps.
}
\label{PM}
\end{figure}

\begin{figure}[ht]
\centering
\includegraphics[width=0.80\linewidth]{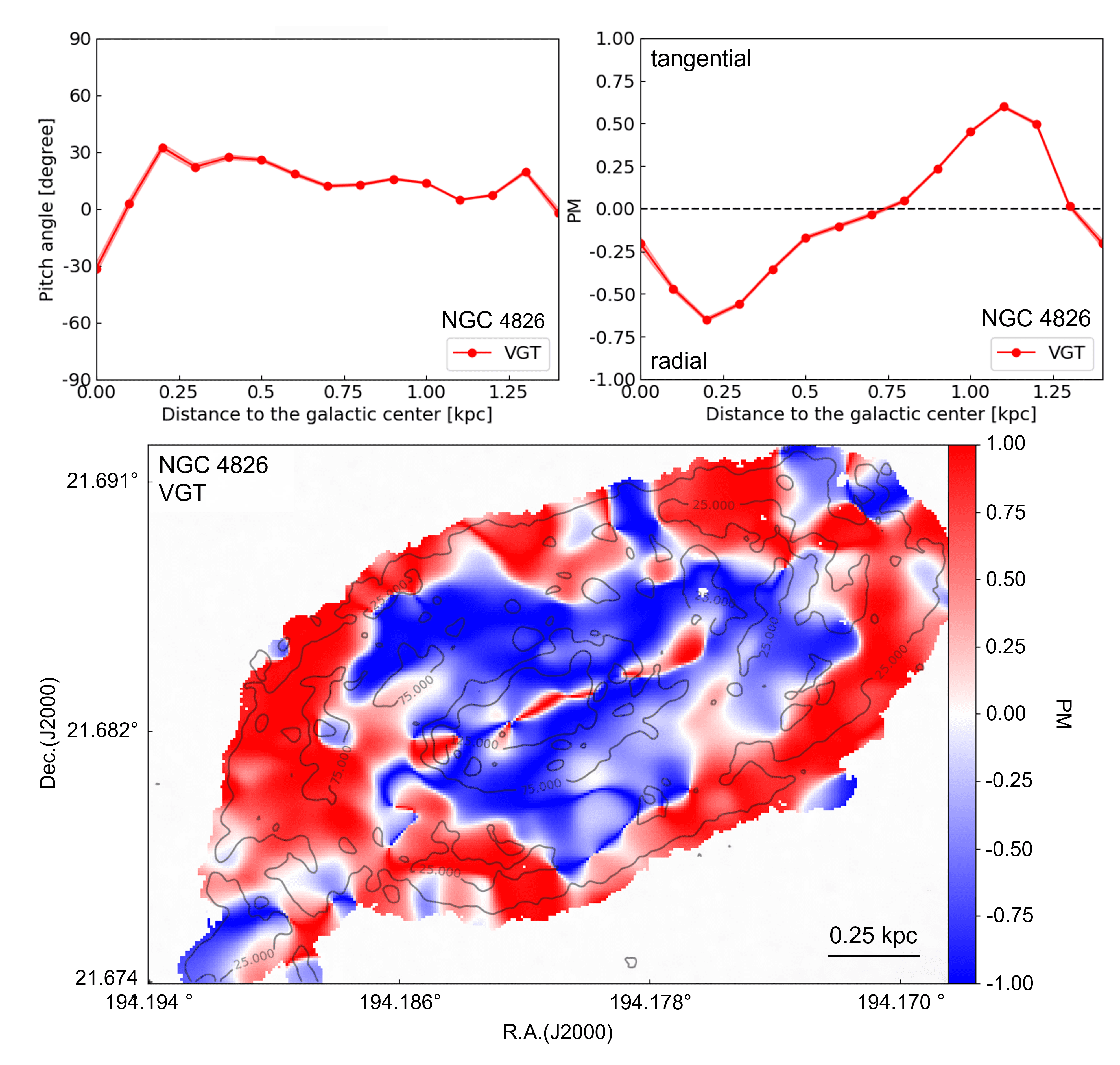}
\caption{Same as Fig.~\ref{AM_rad} and Fig.~\ref{PM}, but for the NGC 4826 galaxy.  The background image and contours are CO emission intensity maps.}
\label{AM_rad2}
\end{figure}

\section{Pitch angle}
\label{sec:pa}
In Figs.~\ref{AM_rad} and \ref{AM_rad2}, we plot the pitch angle as a function of the distance to the galactic center. We follow the recipe used in ref.~\cite{M51} for calculating the pitch angle. We generate a zero pitch angle template by computing the radius $r$ and zero pitch angle, which is perpendicular to the radial direction, of every pixel in galactocentric coordinates. Then we transform the zero pitch angle and radius back to the observer's coordinates, i.e., the POS. The pitch angle is calculated as the difference between the measured position angle of the magnetic field and the template in the IAU convention. The pitch angle is averaged at each annulus from the galaxy center with linearly spaced radial bins. The essential calculation parameters are listed in Tab.~\ref{tab.2}. 

We find M51 has a mean pitch angle around $\sim25^\circ$ and the difference between the three measurements is not significant except at the outskirt $r>5$ kpc region, which is also reported by the ref.\cite{M51}. NGC 1097's three measurements exhibit significantly different pitch angles. Dust polarization appears at a smaller pitch angle. The difference between the VGT and synchrotron polarization at $r\sim1$ kpc is mainly contributed by the contact region of shock front and inner bar, while they have more similarity along the dust lanes. An apparent difference in the VGT and dust polarization in NGC 1068 is observed. Like the ref.\cite{2020ApJ...888...66L}, we find a mean pitch angle $\sim20^\circ$  at $r<1.5$ kpc. Notably, the VGT's pitch angle decreases to $\sim0$ at $r>2.0$ kpc due to the south tail's contribution. NGC 3627 shows a similar pitch angle for both the VGT and dust observation at $r>4.0$ kpc. As for the NGC 4826, a change of pitch angle's sign is observed at $r<0.25$ kpc. 

To quantify the morphology of magnetic field, like the definition of AM, we introduce the Pitch angle Measurement: ${\rm PM}=\cos(2\theta_{p})$, where $\theta_p$ is the pitch angle of the magnetic field. Also, we average the PM at each annulus from the galaxy center. Unlike the averaged pitch angle, PM has a clear physical meaning that $\rm PM>0$ indicates a preferentially tangential field, while the one smaller than 0 represents a preferentially radial field. The difference of PM and pitch angle can be easily understood based on the fact that the angular average of two pitch angles $45^\circ$ and $-45^\circ$ is 0, while these two angles do not have preference being tangential or radial, i.e., $\rm PM = 0$ in this case.

The PM as a function of the distance to the galactic center is presented in Figs.~\ref{AM_rad} and \ref{AM_rad2}. We find that M51, NGC 1068, and NGC 3627 have tangential fields spanning almost all scales. For M51, a negative PM of the VGT at a large distance ($r>5$ kpc ) may come from the noise effect. In addition, NGC 1097 and NGC 4826 exhibit radial fields at small scales ($r<0.5$ kpc for NGC 1097 and $r<0.75$ kpc for NGC 4826) but the tangential field at large scales. In particular, Fig.~\ref{PM} presents the spatial distribution of PM. We find the negative PM usually appears in the positions of inflow. The references used to locate the inner-bars and inflows for Fig.~\ref{PM} and Fig.~\ref{AM_rad2} include: (i) M51: the refs.~\cite{2013ApJ...779...45M,2016A&A...588A..33Q}; (ii) NGC 1097: the refs.~\cite{1995AJ....110..156Q,2009ApJ...702..114D,2010ApJ...723..767V}; (iii) NGC 1068: the refs.~\cite{2014A&A...567A.125G,2020ApJ...888...66L}; (iv) NGC 3627: the refs.~\cite{2009ApJ...692.1623H,2017A&A...597A..85B}. We find the tangential fields, most apparent in the positions of inflows. A similar trend is also observed in NGC 4826, in which the VGT measurement is radial at $r<0.75$ kpc. However, the radial fields cover the entire central region of NGC 4826, instead of only the transition region. 

Moreover, we calculate the AM of the VGT measurement and polarization as a function of the distance to the galactic center. We can see that for M51, the VGT agrees with both synchrotron and dust polarization. However, in NGC 1097, the VGT aligns with synchrotron polarization better than dust polarization. The agreement significantly drops in the inner-bar region. A decreasing trend of AM is observed in NGC 1068 and NGC 3627. The effect of a low signal-to-noise ratio contributes to the misalignment of the VGT and polarization in the south part. A decreasing trend of AM is observed in NGC 1068 and NGC 3627. The effect of a low signal-to-noise ratio, i.e., high VGT uncertainty, may contribute to the misalignment of the VGT and polarization in the south part.

\section{Star formation rate}
\begin{figure}[ht]
\centering
\includegraphics[width=0.99\linewidth]{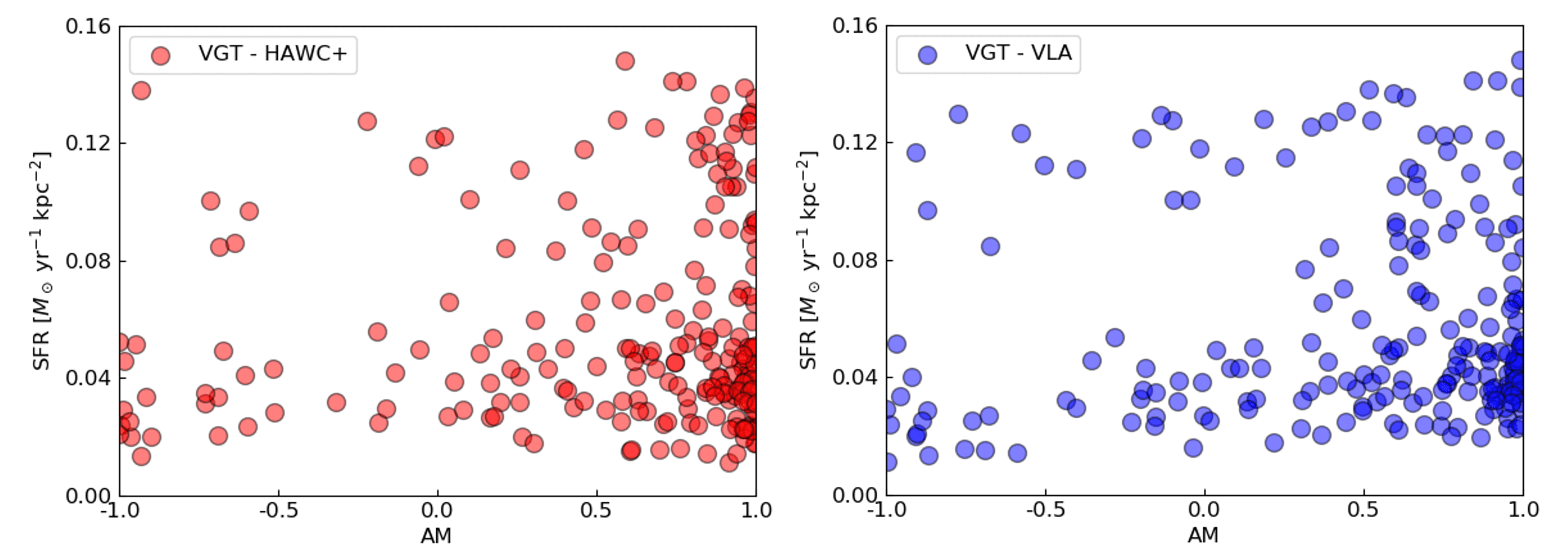}
\caption{The correlation of star formation rate (SFR) and AM for the M51 galaxy. }
\label{SFR}
\end{figure}

In general, the star formation rate (SFR) is not homogeneous across the spiral galaxies. A high SFR means a more active star-forming process, indicating more substantial turbulence in the interstellar medium (ISM) injected by supernovae explosions and winds. Given that the VGT is rooted in MHD turbulence, these mechanisms will generate a relationship between the VGT-traced magnetic field and SFR. Using the available SFR data from the ref.\cite{2019ApJS..244...24L}, we plot the correlation of SFR and AM for the M51 galaxy in Fig.~\ref{SFR}. As expected, high SFR preferentially corresponds to positive AM, which indicates a good agreement of the VGT and polarization. This correspondence is more apparent in dust polarization, which is tightly associated with star formation activity. The correspondence of negative AM and high SFR in synchrotron polarization is contributed by the misalignment observed in the center area of M51.

\end{document}